\documentclass[preprint,prd,aps,showpacs,nofootinbib]{revtex4}
\usepackage{inputenc}
\usepackage{epsfig,float}
\usepackage{amsfonts}
\usepackage[hypertex]{hyperref}
\usepackage{amsmath}
\input{epsf}
\newcommand{\be}{\begin{eqnarray}}
\newcommand{\ee}{\end{eqnarray}}
\newcommand{\ba}{\begin{array}}
\newcommand{\ea}{\end{array}}
\newcommand{\bi}{\begin{itemize}}
\newcommand{\ei}{\end{itemize}}

\restylefloat{figure}
\restylefloat{table}

\begin{document}

\title{Crossed channel analysis of quark and gluon generalized parton distributions with helicity flip}

\author{B.~Pire$^1$,  K.~Semenov-Tian-Shansky$^{2}$, L.~Szymanowski$^{3}$, S.~Wallon$^{4,5}$ }
\affiliation{
$^1$ CPhT, \'{E}cole Polytechnique, CNRS,  91128 Palaiseau, France  \\
$^2$ IFPA, d\'{e}partement AGO,  Universit\'{e} de  Li\`{e}ge, 4000 Li\`{e}ge,  Belgium \\
$^3$ National Centre for Nuclear Research (NCBJ), Warsaw, Poland \\
$^4$ LPT,   Universit\'{e} de Paris-Sud, CNRS, 91405 Orsay, France \\
$^5$ UPMC Universit\'{e} Paris 06, Facult\'{e} de Physique, 4 place Jussieu, 75252 Paris, France
}

\preprint{CPHT-RR009.0314, LPT-Orsay-14-12}

\begin{abstract}
Quark and gluon helicity flip
generalized parton distributions (GPDs)
address the transversity quark and gluon structure of the nucleon.
In order to construct  a theoretically  consistent parametrization of  these hadronic matrix elements,
we work out the set of combinations of those GPDs suitable for the
${\rm SO}(3)$
partial wave (PW) expansion in the cross-channel. This universal result will help to build up a flexible parametrization
of these important hadronic non-perturbative quantities, using for instance the approaches
based on the conformal PW expansion of GPDs such as the Mellin-Barnes integral or the dual parametrization techniques.
\end{abstract}

\maketitle
\thispagestyle{empty}
\renewcommand{\thesection}{\arabic{section}}
\renewcommand{\thesubsection}{\arabic{subsection}}

\section{Introduction}
\label{Sec_Intro}

The transversity quark and gluon structure of the nucleon is a longstanding challenge to
theoretical and experimental studies
\cite{Barone:2001sp}.
Contrarily to the case of the helicity dependent sector, the quark and gluon parts are well separated within
the transversity sector thanks to the chiral-odd property of transversity quark distributions.
Notorious experimental difficulties prevented us up to now from accessing directly quark
transversity distributions through their golden channel,
inclusive dilepton production with transversely polarized beam and target
\cite{RSPax}.
A promising attempt to extract information on this important hadronic sector is currently investigated
with the help of transverse  momentum dependent distributions (TMDs) (see {\it e.g.}
\cite{MELIS:2013zva,Bacchetta}
and references therein).
An alternative method, which may prove to be fruitful, is the study of exclusive reactions, where transversity dependent
generalized parton distributions (GPDs) enter the factorized amplitude in the generalized Bjorken
regime
\cite{Diehl,Diehl:2001pm}.
Here also, the quark and gluon cases are well separated due to the chiral-oddity
of the quark operator. This feature prevents the quark helicity flip GPDs to contribute  to  photon or meson
leptoproduction amplitude at the leading twist
\cite{Diehl:1998pd}
(see however
Ref.~\cite{Ivanov:2002jj,GoloskokovGoldstein,Goloskokov:2013mba}
to evade this no-go theorem). On the other hand, the gluon helicity flip GPDs do not suffer from any selection rule
and appear at  the leading twist level in many amplitudes, for instance in the deeply virtual Compton
scattering (DVCS)
$O(\alpha_s)$
contribution to the leptoproduction of a real photon.
This contribution can be separated through a harmonic analysis
\cite{Diehl:1997bu},
as  discussed in details in
Ref.~\cite{Belitsky:2000jk}.

Whereas  quark helicity flip GPDs can be parametrized thanks to a double distribution
Ansatz \`a la Radyushkin
\cite{Radyushkin:1998es},
provided an educated guess of the shape and normalization of the transversity PDF is used, as in
\cite{ElB}%
\footnote{Other parametrizations have also been recently proposed
in Refs.~\cite{Goldstein:2013gra,Kopeliovich:2014pea}.},
the absence of a forward limit for helicity flip gluon GPDs makes this procedure impracticable.
However, a possible way to get a consistent parametrization of gluon helicity flip GPDs is
to take advantage of a partial wave expansion in the crossed channel. This is the goal of the present paper.

The paper is organized as follows. In
Sec.\ref{Sec_Preliminaries}
we specify our set of conventions for both quark and gluon helicity flip GPDs.
In Sec.~\ref{Sec_SO(3)exp}
we determine the combinations of quark and gluon helicity flip GPDs suitable for the partial wave expansion
in the cross channel
${\rm SO}(3)$
partial waves
by applying the method elaborated in Sec.~4.2 of 
\cite{Diehl}.
This analysis provides an independent cross check of the selection rules for the cross channel exchange quantum numbers
established in
\cite{Hagler:2004yt,Chen:2004cg}
with the help of the general method of X.~Ji and R.~Lebed
\cite{Ji:2000id}.
For reader's convenience we present a short overview of the latter method in
App.~\ref{Sec_Decomposition}.
The cross channel
${\rm SO}(3)$
partial wave (PW) expansion may be used within the advanced model building strategies such as the
approach by D.~Mueller {\it et al.}
\cite{Kumericki:2007sa}
based on the cross channel partial wave expansion of conformal moments of GPDs. Furthermore, this information is useful within the
dual parametrization approach allowing to construct the double partial wave expansion of GPDs (in the conformal and
${\rm SO}(3)$)
partial waves.   In
Sec.~\ref{Sec_Resonance}
we consider the alternative method to construct the double partial wave expansion for quark
and gluon helicity flip GPDs based on the explicit calculation of the cross channel spin-$J$
resonance exchange contributions. We present the explicit results for the case of
$C=-1$
quark helicity flip GPDs.
For the case of helicity flip GPDs these kinds of analysis was never presented in the literature, to the
best of our knowledge.

\section{Preliminaries}
\label{Sec_Preliminaries}

Throughout this paper we adopt the set of conventions of
Ref.~\cite{Diehl}.
The definition of quark helicity flip GPDs involves the nucleon matrix element of
tensor light-cone operator%
\footnote{We employ the light-cone
gauge
$A  \cdot n \equiv A^+=0$,
so that the gauge link does not appear in the operator.}
\be
\hat{O}_T^{q \,\alpha \beta}(- \lambda n/2, \lambda n/2)= \bar{\Psi}(- \lambda n/2)
i \sigma^{\alpha \beta}
\Psi ( \lambda n/2)
\label{TensorOp}
\ee
contracted with the appropriate projector
\be
 n^\alpha g_\bot^{\beta i} \equiv n^\alpha (g^{\beta i}-n^i \bar{n}^\beta- n^\beta \bar{n}^i),
\label{Prj_T}
\ee
where
$n$
and
$\bar{n}$
are the light-cone vectors
($n^2=\bar{n}^2=0$,
$\bar{n} \cdot n=1$)
and the Latin index
$i=1,\,2$
is reserved for the transverse spatial directions.

To the leading twist accuracy the form factor decomposition of the nucleon matrix element of
the operator
(\ref{TensorOp})
involves
$4$
invariant functions
\cite{Diehl:2001pm}:
\be
&&
\frac{1}{2}
\int
\frac{d \lambda}{2 \pi}
e^{i x P^+  \lambda }
\langle N(p' ) |
\bar{\Psi}(- \lambda n/2)
i \sigma^{+i}
\Psi ( \lambda n/2)
| N(p  ) \rangle \nonumber \\ &&
= \frac{1}{2 P^+}
\bar{U}(p')
\left[
H^q_T i \sigma^{+i}+
\tilde{H}_T^q \frac{P^+ \Delta^i- \Delta^+ P^i}{m^2} \right.
\nonumber \\ &&
\left.
+E_T^q \frac{\gamma^+ \Delta^i- \Delta^+ \gamma^i}{2m}+
\tilde{E}_T^q \frac{\gamma^+ P^i- P^+ \gamma^i}{m}
\right] U(p),
\label{def_GPDs_T}
\ee
where $m$ denotes the nucleon mass.
Throughout this paper we  employ the usual kinematical notations for the average momentum
$P$,
$t$-channel momentum transfer
$\Delta$
and the skewness variable
$\xi$:
\be
P= \frac{1}{2}(p+p'); \ \ \Delta=p'-p; \ \ \xi= -\frac{(p'-p) \cdot n}{(p'+p) \cdot n} \equiv -\frac{\Delta^+}{2 P^+}
\label{Kin_Quant_direct}
\ee
defined within the usual DVCS kinematics%
\footnote{We refer to the usual DVCS kinematics
$N(p)+ \gamma^*(q) \rightarrow N(p')+\gamma(q')$:
$-q^2 \equiv Q^2 \rightarrow \infty$, $p \cdot q \rightarrow \infty$,
with fixed
$x_{Bj}=\frac{Q^2}{2 p \cdot q}$
and small negative
$t$.}.
The convolution with the  projecting operator
(\ref{Prj_T})
is implied in
(\ref{def_GPDs_T}).
Each of the
four
invariant functions
$H^q_T$,
$\tilde{H}_T^q$,
$E^q_T$
and
$\tilde{E}_T^q$
depend on the variable
$x$,
skewness
$\xi$,
momentum transfer squared
$\Delta^2 \equiv t$,
as well as on the factorization scale
$\mu$.
Due to  hermiticity and  time reversal invariance, the
four
invariant functions are real valued. Moreover, one may check
\cite{Diehl:2001pm}
that
$H^q_T$, $\tilde{H}_T^q$, $E^q_T$
are even functions of
$\xi$
while
$\tilde{E}_T^q$
is an odd function of
$\xi$.

For consistency we provide the relation of the parametrization
(\ref{def_GPDs_T})
to that used by Z.~Chen and X.~Ji
\cite{Chen:2004cg},
which reads
\be
&&
H^q_T|_{{\rm eq.} \; (3)}= H_{Tq} |_{{\rm ref.} \; [19]
};
\ \ \
E^q_T|_{{\rm eq.} \; (3)}= -E_{Tq} |_{{\rm ref.} \; [19]
}; \nonumber  \\ &&
\tilde{H}^q_T|_{{\rm eq.} \; (3)}= -\frac{1}{2} \tilde{H}_{Tq} |_{{\rm ref.} \; [19]
};
\ \ \
\tilde{E}^q_T|_{{\rm eq.} \; (3)}= -\frac{1}{2} \tilde{E}_{Tq} |_{{\rm ref.} \; [19]
}.
\ee

In the definition of quark helicity flip GPDs instead of the tensor current operator
(\ref{TensorOp})
one can use the pseudotensor current employing
the relation
\cite{Diehl:2001pm}%
\footnote{Throughout this paper we use the conventions employed in
\cite{Diehl},
\cite{Diehl:2001pm}:
$\sigma^{\mu \nu}=\frac{i}{2} [\gamma^\mu,\,\gamma^\nu] $,
$\gamma_5=i \gamma^0 \gamma^1 \gamma^2 \gamma^3$
and the Levi-Civita tensor defined with
$\varepsilon_{0123}=+1$.
It worths mentioning that this definition of the Levi-Civita tensor differs from that employed {\it e.g.}
in \cite{IZ}, where
$\varepsilon^{0123}=+1$.
 }:
\be
\sigma^{\alpha \beta} \gamma_5= - \frac{i}{2} \varepsilon^{\alpha \beta \gamma \delta} \sigma_{\gamma \delta}.
\ee
Therefore, the relation between the tensor and pseudotensor currents reads
\be
\hat{\tilde{O}}_T^{q\,\alpha \beta} (-\lambda n /2 ,\lambda n /2) \equiv  \bar{\Psi}(-\lambda n /2)
 \sigma^{\alpha \beta} \gamma_5 \Psi (\lambda n /2) =- \frac{i}{2} \varepsilon^{\alpha \beta \gamma \delta} \hat{O}_{T \, \gamma \delta}^q(-\lambda n /2 ,\lambda n /2).
 \label{PseudoTensorOp}
\ee
As a result, the equivalent parametrization for the quark helicity flip GPDs from the Fourier transform of
the pseudotensor operator reads
\be
&&
\frac{1}{2} \int \frac{d \lambda}{2 \pi} e^{i x P^+ \lambda} \langle p' | \bar{\Psi}(- \lambda n /2) \sigma^{+ i} \gamma_5 \Psi(\lambda n /2)| p \rangle
\nonumber \\ &&
= - \frac{1}{2} \varepsilon^{+ i \gamma \delta}
\frac{1}{2} \int \frac{d \lambda}{2 \pi} e^{i x P^+ \lambda} \langle p' | \bar{\Psi}(-\lambda n /2)i  \sigma_{\gamma \delta}   \Psi(\lambda n /2)| p \rangle
\nonumber \\ &&
=- \frac{1}{2}  \varepsilon^{+ i \gamma \delta} \bar{U}(p') \Big[ H_T^q i \sigma_{\gamma \delta}
+ \tilde{H}_T^q \frac{P_\gamma \Delta_\delta-\Delta_\gamma P_\delta}{m^2}
+E_T^q  \frac{\gamma_\gamma \Delta_\delta-\Delta_\gamma \gamma_\delta}{2m}
+\tilde{E}_T^q  \frac{\gamma_\gamma P_\delta-P_\gamma \gamma_\delta}{m}
\Big]U(p)
\nonumber \\ &&
=\bar{U}(p') \Big[ H_T^q \sigma^{+ i} \gamma_5 -\tilde{H}_T^q \frac{\varepsilon^{+i P \Delta}}{m^2}
-E_T^q \frac{\varepsilon^{+i \gamma \Delta} \gamma_\gamma}{2m}
-\tilde{E}_T^q \frac{\varepsilon^{+i \gamma P} \gamma_\gamma}{m}
\Big]U(p),
\label{Set_of_Ptensor}
\ee
where in the last line we use the abbreviate notations for the contraction of the Levi-Civita tensor
with four-vectors.


Let us  now turn to gluon helicity flip GPD defined from the nucleon matrix element of
the appropriate projection of the  gluon light-cone operator:
\be
\hat{O}_T^{g \,\alpha \rho \, \beta \sigma} (- \lambda n/2, \lambda n/2) = G^{\alpha \rho}(- \lambda n/2) G^{\beta \sigma}( \lambda n/2).
\label{op_free_index}
\ee
The corresponding projection operation reads
\be
{\mathbb{S}} G^{+ i} (- \lambda n/2) G^{j + } ( \lambda n/2) \equiv
\tau_{i j; \, \rho \sigma}^\bot n^\alpha n^\beta G^{\alpha \rho}(- \lambda n/2) G^{\beta \sigma}( \lambda n/2),
\label{op_gluon_hel_flip}
\ee
where the
$\mathbb{S}$
symbol stands for the symmetrization in the two transverse spatial indices and removal of the corresponding trace.
The explicit
expression for the
$\mathbb{S}$
operation has the form of the following projecting operator
\cite{Belitsky:2000jk}:
\be
\tau^{\bot}_{ij ; \, \rho \sigma}=\frac{1}{2} \left( g^{\bot}_{i \rho} g^{\bot}_{j \sigma} +g^{\bot}_{j \rho} g^{\bot}_{i \sigma} -
g^{\bot}_{i j} g^{\bot}_{\rho \sigma}   \right),
\label{projection}
\ee
where
$
g^{\bot}_{\rho \sigma}= g_{\rho \sigma}-n_\rho \bar{n}_\sigma- n_\sigma \bar{n}_\rho
$.

The parametrization of the nucleon matrix element of the gluon helicity flip operator
(\ref{op_gluon_hel_flip})
to the leading twist accuracy involves
four
invariant functions
\cite{Diehl}:
\be
&&
\frac{1}{P^+} \int \frac{d \lambda }{2 \pi} e^{i x P^+ \lambda}
\langle p'|  {\mathbb{S}} G^{+ i} (- \lambda n/2) G^{j + } ( \lambda n/2) | p \rangle 
\nonumber
\\ &&
= {\mathbb{S}} \; \frac{1}{2 P^+} \frac{P^+ \Delta^j-\Delta^+ P^j}{2 m P^+} \bar{U}(p') \Big[ H_T^g i \sigma^{+i}+ \tilde{H}_T^g  \frac{P^+ \Delta^i-\Delta^+ P^i}{m^2}
\nonumber
\\ &&
+E_T^g  \frac{\gamma^+ \Delta^i-\Delta^+ \gamma^i}{2m}+
\tilde{E}_T^g  \frac{\gamma^+ P^i-P^+ \gamma^i}{m} \Big] U(p),
\label{Diehl_def}
\ee
with
$H_T^g$,
$E_T^g$,
$\tilde{H}_T^g$
and
$\tilde{E}_T^g$
being  functions of the usual GPD variables.
Similarly to the quark case, from the combination of hermiticity and
$T$-invariance they are real valued.
$H_T^g$, $E_T^g$, $\tilde{H}_T^g$
are even functions of
$\xi$
while
$\tilde{E}_T^g$
is an odd function of
$\xi$.
Moreover, from the $C$-invariance
$H_T^g$, $E_T^g$, $\tilde{H}_T^g$
and
$\tilde{E}_T^g$
are shown to be even functions of
$x$.

Again let us specify the relation of the parametrization
(\ref{Diehl_def})
to that used by Chen and Ji
\cite{Chen:2004cg}:
\be
&&
H^g_T|_{{\rm eq.} \; (12)}= -2x H_{Tg} |_{{\rm ref.} \; [19]
};
\ \ \
E^g_T|_{{\rm eq.} \; (12)}= -2 x E_{Tg} |_{{\rm ref.} \; [19]
}; \nonumber  \\ &&
\tilde{H}^g_T|_{{\rm eq.} \; (12)}= -x \tilde{H}_{Tg} |_{{\rm ref.} \; [19]
};
\ \ \
\tilde{E}^g_T|_{{\rm eq.} \; (12)}= -x \tilde{E}_{Tg} |_{{\rm ref.} \; [19]
}.
\ee
Note, that the pioneering papers
\cite{Hoodbhoy:1998vm},
\cite{Belitsky:2000jk}
overlooked the distributions
$\tilde{H}_T^g$
and
$\tilde{E}_T^g$
(see the discussion in
\cite{Diehl:2001pm}
and
\cite{Chen:2004cg}).

Similarly to the case of quark helicity flip GPDs (which possess two equivalent definitions from the
nucleon matrix elements of tensor and pseudotensor quark operators) there is an equivalent
definition of gluon helicity flip GPDs involving the dual gluon field strength.
Indeed, the gluon non-local operator
(\ref{op_free_index})
has the following relation to the dual gluon non-local operator:
\be
\hat{\tilde{O}}_T^{g \,\alpha \rho \, \beta \sigma} (- \lambda n/2, \lambda n/2) =
\tilde{G}^{\alpha \rho}(-\lambda n/2) G^{\beta \sigma}( \lambda n/2)= \frac{1}{2}
 \varepsilon^{\alpha \rho \gamma \tau} G^{\gamma \tau}(- \lambda n/2)  G^{\beta \sigma}( \lambda n/2).
\ee
This leads to the equivalent parametrization of gluon helicity flip GPDs:
\be
&&
\frac{1}{P^+} \int \frac{d \lambda}{2 \pi} e^{i x P^+ \lambda}
\langle p'| \, {\mathbb{S}} 
\tilde{G}^{+ i} (-\lambda n/2 ) G^{j + } (\lambda n/2) | p \rangle 
\nonumber
\\ &&
= {\mathbb{S}} \; \frac{1}{2 P^+} \frac{P^+ \Delta^j-\Delta^+ P^j}{2 m P^+} \bar{U}(p') \Big[ -H_T^g   \sigma^{+i} \gamma_5+ \tilde{H}_T^g  \frac{ \varepsilon^{+i P \Delta} }{m^2}
\nonumber
\\ &&
+E_T^g  \frac{\varepsilon^{+i \gamma \Delta} \gamma_\gamma   }{2m}+
\tilde{E}_T^g  \frac{\varepsilon^{+i \gamma P}\gamma_\gamma  }{m} \Big] U(p).
\label{def_helicity_flip_gluons_equivalent}
\ee
The presence of the equivalent definition
(\ref{def_helicity_flip_gluons_equivalent})
just mirrors the fact that, exactly as the quark helicity flip operator,
the gluon helicity flip operator
(\ref{op_gluon_hel_flip})
does not possess a definite
$P$
parity. We review this issue  in
Appendix~\ref{Sec_Decomposition}
in which the quantum number selection rules for the cross channel exchanges contributing to the Mellin moments
of helicity flip GPDs are considered.   Surprisingly, to the best of our knowledge, the definition
(\ref{def_helicity_flip_gluons_equivalent})
 was never previously discussed in the literature.
We take advantage of the existence of two equivalent parametrizations for both quark and gluon helicity flip
GPDs 
in Sec.~\ref{Sec_Resonance}
when   considering the contributions of the cross-channel spin-$J$ resonance exchanges of natural 
($P=(-1)^J$) 
and  unnatural   
($P=(-1)^{J+1}$)
parity.

\section{${\rm SO}(3)$ partial wave decomposition of   quark and gluon GPDs with helicity flip}
\label{Sec_SO(3)exp}

Building up the phenomenological Ans\"{a}zte for GPDs
in consistency with the fundamental theoretical requirements
(such as the polynomiality, analyticity, positivity, Regge theory constraints {\it etc.})
is eagerly awaited by the present day phenomenology but
represents a considerable theoretical challenge.
Historically, the first successful parametrization of GPDs was based on the spectral
representation of GPDs in terms of double distributions
\cite{RDDA1,RDDA2,RDDA3,RDDA4},
which is the most straightforward way
to implement the polynomiality property of GPDs.
The alternative way to proceed relies on the expansion of GPDs over the convenient systems of orthogonal polynomials
in order to achieve the factorization of functional dependencies of GPDs on their variables.
As the first step for such expansion one usually employs the set of eigenfunctions of the leading order (LO) evolution equations
which leads to the expansion of GPDs over the basis of the conformal partial waves
\cite{Mueller:2005ed}.

One of the ways to proceed with the conformal partial wave expansion of
GPDs is to further expand the conformal moments over a basis of suitable orthogonal polynomials carrying the labels of irreducible 
representations of the cross channel%
\footnote{The term ``cross channel''  refers to the $t$-channel of the DVCS reaction: $ \gamma^*(q)+\gamma(-q') \to N(p') + \bar{N}(-p)$.}
angular momentum
${\rm SO}(3)$
rotation group
\cite{Polyakov:1998ze,Polyakov:2002wz,Kumericki:2007sa}.
In this way one arrives to a double partial wave expansion of GPDs (both over the conformal basis and in the cross channel 
partial waves). Different methods were proposed in the literature to handle the double partial wave expansions of GPDs
(for the discussion see 
{\it e.g.}
\cite{Mueller:2005ed}).

One of such methods is the framework of the so-called dual parametrization of GPDs
\cite{Polyakov:2002wz,Polyakov:2008aa}.
Within this approach the operator matrix elements defining GPDs are seen as  infinite sums of the cross channel
resonance exchanges of arbitrary high
spin
$J$.
The double partial wave expansion is first assigned meaning in the cross channel, where it rather represents 
generalized distribution amplitude (GDA). Then, exploiting the crossing symmetry, it is analytically
continued to the direct channel allowing to work out a rigorous expression for GPDs. The term ``dual''  
emphasizes  the natural association with the old idea of duality in hadron-hadron low-energy scattering, 
that for binary scattering can roughly be summarized as the assumption that the infinite sum over only just  
the cross-channel Regge exchanges may provide the complete description of the process within certain 
kinematical domain
\cite{GreenSchwarzWitten}.
The Ansatz for the
${\rm SO}(3)$
partial wave amplitudes was also suggested within the Mellin-Barnes transform technique developed in
\cite{Kumericki:2007sa}.
Although it employs rather different mathematical tools, it should be in general equivalent to the 
dual parametrization approach.

However, even without any respect to a summation method employed to handle the double partial wave expansions,
finding out the combinations of GPDs suitable for the $t$-channel
${\rm SO}(3)$
partial waves and the choice of the appropriate basis of the orthogonal polynomials represents an important task.
For example, for the case of the unpolarized quark and gluon nucleon GPDs this kind of analysis gives rise to the
so-called electric and magnetic combinations of GPDs 
\cite{Diehl}:
\be
H^{E \, \{q,g\}}=H^{\{q,g\}}+ \tau E^{\{q,g\}}; \ \ \
H^{M \,\{q,g\}}=H^{\{q,g\}}+ E^{\{q,g\}},
\ee
where
\be
\tau \equiv \frac{\Delta^2}{4 m^2}.
\ee
These combinations are to be expanded respectively in terms of
$ P_J(\cos \theta)
$
and
$ P'_J(\cos \theta)$,  
where
$P_J(\chi)$
stand for the Legendre polynomials and
$\theta$
refers for the
$t$-channel scattering angle in the
$N \bar{N}$
center-of-mass frame. The resulting double partial wave expansion for
the electric and magnetic combinations of unpolarized and polarized nucleon quark and gluon GPDs within the dual 
parametrization approach was presented in
\cite{Polyakov:2008aa}
and
\cite{SemenovTianShansky:2010zv}.

In this section we address the problem of pointing out the combinations of quark and gluon helicity flip GPDs
suitable for the expansion in the $t$-channel
${\rm SO}(3)$ partial waves. We employ the method
suggested in Sec.~4.2 of
\cite{Diehl}.
In order to identify  the combinations of quark GPDs with helicity flip suitable for the partial 
wave expansion in the $t$-channel partial waves one has to consider the analytically continued to the 
cross channel form factor decomposition of $N$-th Mellin moments of the corresponding operator matrix elements. 
The  analytically continued matrix Mellin moments are then computed in a specific reference frame
($N \bar{N}$
center-of-mass) for definite (usual)
helicity of nucleons (here denoted as
$\lambda$ and $\lambda'$).
This allows to specify the rotational functions
$d^J_{J_3, \, |\lambda-\lambda'|}$
governing the polar angle dependence.
Also this methods provides a cross check of the
$J^{PC}$
quantum number selection rules worked out in
Refs.~\cite{Hagler:2004yt,Chen:2004cg}
by the method of X.~Ji and R.~Lebed
\cite{Ji:2000id} (see App.~\ref{Sec_Decomposition}
for a review).

\subsection{${\rm SO}(3)$ partial wave decomposition of quark helicity flip GPDs}
\label{SubSec_PWquark}

Following the receipt of Sec. 4.2 of 
Ref.~\cite{Diehl},  
in order to identify  the combinations of quark helicity flip  GPDs suitable
for the partial wave expansion in the $t$-channel partial waves we
consider the form factor decomposition of the $N$-th Mellin moments
(\ref{FF_dec_Haegler})
of quark helicity flip  GPDs analytically continued the to the cross channel
($t>0$).
Thus we are dealing with the form factor decomposition of
$N$-th
Mellin moments  of  quark helicity flip
$N \bar{N}$
GDAs.

We establish the following notations for the kinematical quantities
(\ref{Kin_Quant_direct})
analytically continued to the cross channel:
\be
&&
t \equiv \Delta^2 \rightarrow \tilde{s}; \nonumber \\ &&
\Delta \equiv p'-p \rightarrow  \tilde{P} \equiv p'+\tilde{p}; \nonumber \\ &&
P= \frac{p'+p}{2} \rightarrow \frac{1}{2} \tilde{\Delta} \equiv \frac{p'-\tilde{p}}{2}.
\label{Kin_Quant_cross}
\ee

The form factor decomposition of the $N$-th Mellin moment of the quark helicity flip
$N \bar{N}$
GDA then reads:
\be
&&
 { \mathbb{S}}_{\{ \nu \mu_1...\mu_N\}}   
 \langle  N(p',\lambda') \bar{N}( \tilde{p},\lambda) | \bar{\Psi}(0) i \sigma^{\mu \nu} (\overleftrightarrow{i D}_{\mu_1})...(\overleftrightarrow{i D}_{\mu_N}) \Psi(0)| 0 \rangle
 \nonumber \\ &&
 = { \mathbb{S}}_{\{ \nu \mu_1...\mu_N\}} \; \bar{U}(p',\lambda') \Big[ \sum_{k=0 \atop {\rm even}}^N
 \Big\{ i \sigma^{\mu \nu} \tilde{P}^{\mu_1} \ldots \tilde{P}^{\mu_k}  \frac{1}{2} \tilde{\Delta}^{\mu_{k+1}} \ldots \frac{1}{2} \tilde{\Delta}^{\mu_{N}} A^q_{T \, N+1,k}(\tilde{s})  \nonumber \\ &&
 +\frac{ \frac{1}{2} \tilde{\Delta}^\mu \tilde{P}^\nu-\frac{1}{2} \tilde{\Delta}^\nu \tilde{P}^\mu}{m^2}  \tilde{P}^{\mu_1} \ldots \tilde{P}^{\mu_k}  \frac{1}{2} \tilde{\Delta}^{\mu_{k+1}} \ldots \frac{1}{2} \tilde{\Delta}^{\mu_{N}} \tilde{A}^q_{T \, N+1,k}(\tilde{s})
  \nonumber \\ &&
 +  \frac{\gamma^\mu \tilde{P}^\nu- \gamma^\nu \tilde{P}^\mu}{m^2}  \tilde{P}^{\mu_1} \ldots \tilde{P}^{\mu_k}  \frac{1}{2} \tilde{\Delta}^{\mu_{k+1}} \ldots \frac{1}{2} \tilde{\Delta}^{\mu_{N}} B^q_{T \, N+1,k}(\tilde{s}) \Big\}
   \nonumber \\ &&
+ \sum_{k=0 \atop {\rm odd}}^N
 \frac{\gamma^\mu \frac{1}{2} \tilde{\Delta}^\nu- \gamma^\nu \frac{1}{2} \tilde{\Delta}^\mu}{m^2}  \tilde{P}^{\mu_1} \ldots \tilde{P}^{\mu_k}  \frac{1}{2} \tilde{\Delta}^{\mu_{k+1}} \ldots \frac{1}{2} \tilde{\Delta}^{\mu_{N}} \tilde{B}^q_{T \, N+1,k}(\tilde{s})
 \Big] V(\tilde{p},\lambda),
 \label{FF_dec_Haegler_AC}
\ee
where
$U$
and
$V$
are the usual Dirac spinors and
$\lambda'$ ($\lambda$)
denote the corresponding nucleon (antinucleon) helicity.
The generalized form factors
$A^q_{T}$ $\tilde{A}^q_{T}$, $B^q_{T}$, $\tilde{B}^q_{T}$
introduced in
(\ref{FF_dec_Haegler})
are analytically continued to the
cross channel.

Now, in order to find which partial waves can contribute into the  matrix elements 
(\ref{FF_dec_Haegler_AC}), 
we compute the corresponding spin-tensor structures for spinors of definite (usual) helicity in the
$N \bar{N}$
center-off-mass (CMS) frame using the explicit expressions
(\ref{helicity_spinors})
for the nucleon spinors  with definite ordinary helicity.

For this issue we introduce the following parametrization of the relevant $3$-vectors in the
$N \bar{N}$
CMS:
\be
&&
\vec{p'}=   \frac{\sqrt{\tilde{s}}}{2} \beta \big\{ \sin \theta \cos \phi, \, \sin \theta \sin \phi, \,  \cos \theta \big\} ; 
\nonumber \\ &&
\vec{\tilde{p}}=  \frac{\sqrt{\tilde{s}}}{2} \beta   \big\{  \sin (\pi -\theta) \cos (\phi+\pi) ,\sin (\pi -\theta) 
\sin (\phi+\pi), \cos (\pi -\theta) \big\},
\ee
where
$\theta \in [0, \pi]$
and
$\phi \in [0,2\pi]$
are the usual polar and azimuthal angles and 
$\beta$ 
denotes the relativistic velocity
$\beta= \sqrt{1- \frac{4m^2}{\tilde{s}}}$.
The four-vectors
$\tilde{P}$
and
$\tilde{\Delta}$
(\ref{Kin_Quant_cross})
in the
$N \bar{N}$
CMS then read:
\be
&&
\tilde{P} \equiv p'+\tilde{p}= ( \sqrt{\tilde{s}}, 0,0,0); \nonumber \\ &&
\tilde{\Delta} \equiv p'-\tilde{p}=  \sqrt{\tilde{s}}   \beta (0, \sin \theta \cos \phi, \, \sin \theta \sin \phi, \,  \cos \theta).
\ee

We are now about to compute the matrix elements for the $N$-th Mellin moments of 
$N \bar{N}$ 
quark helicity flip GDA
\be
\langle  N(p',\lambda') \bar{N}( \tilde{p},\lambda)| \hat{O}_{ T}^{q\, + i, \, ++ \ldots + }|0 \rangle
\label{NthMellinGDA}
\ee
in the
$\bar{N} N$ CMS
for the cases when the helicities of nucleon and antinucleon couple to
$\lambda'-\lambda=0$
and
$|\lambda'-\lambda|=1$.

We project out the combination of
(\ref{NthMellinGDA})
with definite helicity
$J_3=\pm 1$ of the corresponding operator:
\be
&&
\langle  N(p',\lambda') \bar{N}( \tilde{p},\lambda)| \hat{O}_{ T}^{q \, + 1, \, ++ \ldots + }|0 \rangle  
\pm i \langle  N(p',\lambda') \bar{N}( \tilde{p},\lambda)| \hat{O}_{  T}^{q\, + 2, \, ++ \ldots + }|0\rangle
\nonumber \\ &&
\equiv \langle  N(p',\lambda') \bar{N}( \tilde{p},\lambda)| \hat{O}_{  T}^{q \, + (1 \pm i2), \, ++ \ldots + }|0 \rangle.
\label{helicity_combinations_quarks}
\ee

Then for
$\lambda= \lambda'$
{\it i.e.} the aligned configuration of nucleon and antinucleon helicities%
\footnote{Note, that throughout this section the hadron helicity labeling refers to the $t$-channel. Obviously, 
when crossing back to the direct channel the helicity
$\lambda$
is reversed. }
($N^\uparrow \bar{N}^\uparrow$ or $N^\downarrow \bar{N}^\downarrow$)
we get
\be
&&
\left.\langle  N(p',\lambda') \bar{N}( \tilde{p},\lambda)| \hat{O}_{  T}^{q\, + (1+i2), \, ++ \ldots + }|0 \rangle 
\right|_{\lambda=\lambda'} 
= \eta_{\lambda' \lambda}^+ (\tilde{P}^+)^{N+1} \sin \theta  \nonumber \\ && \times \sum_{k=0 \atop {\rm even}}^N
\big[ (\beta-1) A_{T \, N+1,k}^q(\tilde{s}) + \beta^2 \frac{\tilde{s}}{2 m^2}  \tilde{A}_{T \, N+1,k}^q(\tilde{s}) -B_{T \, N+1,k}^q(\tilde{s})
\big] (\frac{1}{2} \beta \cos \theta)^{N-k};
\label{Result_LequalLp_plus}
\ee
and
\be
&&
\left. \langle  N(p',\lambda') \bar{N}( \tilde{p},\lambda)| \hat{O}_{ T}^{q\,+ (1-i2), \, ++ \ldots + }|0 \rangle \right|_{\lambda=\lambda'}  
= \eta_{\lambda' \lambda}^- (\tilde{P}^+)^{N+1} \sin \theta  \nonumber \\ && \times \sum_{k=0 \atop {\rm even}}^N
\big[ -(\beta+1) A_{T \, N+1,k}^q(\tilde{s}) + \beta^2 \frac{\tilde{s}}{2 m^2}  \tilde{A}_{T \, N+1,k}^q(\tilde{s}) -B_{T \, N+1,k}^q(\tilde{s})
\big] (\frac{1}{2} \beta \cos \theta)^{N-k}.
\label{Result_LequalLp_minus}
\ee
Note, that the combinations
(\ref{helicity_combinations_quarks})
possess definite phases depending on the azimuthal angle
$\phi$
denoted as
$\eta_{\lambda' \lambda}^\pm$.
Now one can decompose
(\ref{Result_LequalLp_plus})
and
(\ref{Result_LequalLp_minus})
in the partial waves with total angular momentum
$J$.
The
$\theta$
dependence is governed by the Wigner ``small-$d$'' rotation functions
 $d^J_{J^3, |\lambda'-\lambda|}$.
In this case we have
$|\lambda'-\lambda|=0$
and
$J_3= \pm 1$. Therefore, one has to use
\be
d^J_{\pm 1, 0}(\theta) = (\pm 1) \frac{1}{\sqrt{J(J+1)} } \sin \theta \, P'_J(\cos \theta).
\ee

After the inverse crossing
(\ref{Kin_Quant_cross})
back to the $s$-channel, within the DVCS kinematics
$\cos \theta$
up to higher twist corrections becomes
\be
\cos \theta \rightarrow \frac{1}{\xi \beta}+ O(1/Q^2).
\ee
At this stage we switch to massless hadrons so that we could consider
hadron helicities as true quantum numbers thus making simple
the crossing relation between the corresponding partial amplitudes (in particular excluding mixing).
This implies setting
$\beta=1$
(which means systematically neglecting the threshold corrections
$\sim \sqrt{1-\frac{4m^2}{t}}$).
However, up to the very end  we keep the non-zero mass within the Dirac spinors.
It would be fair to say that this step is somewhat cumbersome and requires further study.
A possible solution to the problem of threshold singularities could be the appropriate resummation
of the cross channel partial wave expansion in order to avoid the appearing of the kinematical
singularities in the direct channel. Some attempts to follow this program were performed within
the dual parametrization approach 
\cite{SemenovPhD}.

Putting aside the mentioned above problem,
we conclude that the following combinations of quark helicity flip GPDs are to be expanded in
$P'_J(1/\xi)$:
\be
&&
\tau
\tilde{H}_T^q(x,\xi,\Delta^2) -\frac{1}{2} E_T^q(x,\xi,\Delta^2); \nonumber \\ &&
-  H_T^q(x,\xi,\Delta^2)+
\tau
\tilde{H}_T^q(x,\xi,\Delta^2) -\frac{1}{2} E_T^q(x,\xi,\Delta^2).
\label{combi}
\ee

Now we consider the case when
$|\lambda'-\lambda|=1$
({\it i.e.} the opposite  helicity configuration of nucleon and antinucleon:
$N^\uparrow \bar{N}^\downarrow$ or $N^\downarrow \bar{N}^\uparrow$).
For the operator helicity
$J_3= \pm 1$
configurations
(\ref{helicity_combinations_quarks})
we get
\be
&&
\left. \langle  N(p',\lambda') \bar{N}( \tilde{p},\lambda)| \hat{O}_{  T}^{q\,+ (1+i2), \, ++ \ldots + }|0 \rangle \right|_{|\lambda'-\lambda|=1} 
\nonumber \\ &&
= {\eta}_{\lambda' \lambda}^+ (\tilde{P}^+)^{N+1} (1+\cos \theta)    \Big\{  \sum_{k=0 \atop {\rm even}}^N
\big[  \frac{2m}{\sqrt{\tilde{s}}} A_{T \, N+1,k}^q(\tilde{s}) +   \frac{\sqrt{\tilde{s}}}{2 m}  B_{T \, N+1,k}^q(\tilde{s})
\big] (\frac{1}{2} \beta \cos \theta)^{N-k}
\nonumber \\ &&
-\sum_{k=0 \atop {\rm odd}}^N
\frac{\beta \sqrt{\tilde{s}}}{2m}  \tilde{B}_{T \, N+1,k}^q (\tilde{s}) (\frac{1}{2} \beta \cos \theta)^{N-k} \Big\}
\label{Result_L_Non_equalLp_plus}
\ee
and
\be
&&
\left. \langle  N(p',\lambda') \bar{N}( \tilde{p},\lambda)| \hat{O}_{  T}^{q\, + (1-i2), \, ++ \ldots + }|0 \rangle
 \right|_{|\lambda'-\lambda|=1}
\nonumber \\ &&
= {\eta}_{\lambda' \lambda}^- (\tilde{P}^+)^{N+1} (1-\cos \theta)    \Big\{  \sum_{k=0 \atop {\rm even}}^N
\big[ \frac{2m}{\sqrt{\tilde{s}}} A_{T \, N+1,k}^q(\tilde{s}) +   \frac{\sqrt{\tilde{s}}}{2 m}  B_{T \, N+1,k}^q(\tilde{s})
\big] (\frac{1}{2} \beta \cos \theta)^{N-k}
\nonumber \\ &&
+\sum_{k=0 \atop {\rm odd}}^N
\frac{\beta \sqrt{\tilde{s}}}{2m}  \tilde{B}_{T \, N+1,k}^q(\tilde{s}) (\frac{1}{2} \beta \cos \theta)^{N-k} \Big\},
\label{Result_L_Non_equalLp_minus}
\ee
where 
$ {\eta}_{\lambda' \lambda}^\pm$ 
denote the azimuthal angle 
$\phi$ 
dependent phases.

The combinations
(\ref{Result_L_Non_equalLp_plus}),
(\ref{Result_L_Non_equalLp_minus})
are to be expanded respectively in
\be
d^J_{1, 1}(\theta)=   \frac{1}{J(J+1)} (1+\cos \theta) \big[ P'_J(\cos \theta) + \cos \theta P''_J(\cos \theta) - P''_J(\cos \theta) \big]
\ee
and
\be
d^J_{-1, 1}(\theta)=  \frac{1}{J(J+1)} (1-\cos \theta) \big[  P'_J(\cos \theta) + \cos \theta P''_J(\cos \theta) + P''_J(\cos \theta) \big].
\ee

Performing crossing to the direct channel we conclude that the combinations of quark helicity flip GPDs
\be
H_T^q(x,\xi,\Delta^2)+ 
\tau \tilde{H}_T^q(x,\xi,\Delta^2)  \pm
\tau \tilde{E}_T^q(x,\xi,\Delta^2)
\label{comb_without_definite}
\ee
are to be expanded in
\be
 P'_J(1/\xi) +   \frac{1 \mp \xi}{\xi}  P''_J(1/\xi).
\ee

Comparing
(\ref{Result_LequalLp_plus}),
(\ref{Result_LequalLp_minus}),
(\ref{Result_L_Non_equalLp_plus}),
(\ref{Result_L_Non_equalLp_minus})
with the parametrization for the $N$-th Mellin moments of
quark helicity flip GPDs 
(\ref{FF_dec_Haegler})
we work out the set of the selection rules for the
$J^{PC}$
quantum numbers for the $t$-channel resonance exchanges
contributing into $N$-th Mellin moments of quark helicity flip GPDs.
Theses selection rules coincide with those
worked out with the method of X.~Ji and R.~Lebed reviewed in
Appendix~{\ref{Sec_Decomposition}}.

As the example, let us consider the
Mellin moments of combinations
(\ref{combi})
of quark helicitity flip
GPDs.
Note, that  $T$-invariance constraints are implemented directly through
the requirement that the Mellin moments of
$H_T^q$,
$\tilde{H}_T^q$
and
$E_T^q$
should contain only even powers of $\xi$, while those of
$\tilde{E}_T^q$
involve only odd powers of
$\xi$.

For definiteness, let us consider the even Mellin moments
($N=0,\,2,\, \ldots$)
of
(\ref{combi}).
Therefore, we are dealing with non-singlet 
({\it i.e.} 
$C=-1$)
combinations%
\footnote{In particular, the non-singlet combination of $H_T^q$ in the forward limit $\xi=0$, $\Delta^2=0$
is reduced to 
$H_T^{q-}(x,0,0)=\delta q(x)- \delta \bar{q} (x)$, 
where 
$\delta q(x)$ 
is the quark transversity distribution.
}.
Following our analysis,
the matrix elements
(\ref{Result_LequalLp_plus})
and
(\ref{Result_LequalLp_minus})
are to be expanded respectively in
$d^J_{\pm 1, 0}(\theta)$.
From  $T$-invariance requirements the summation over
$k$
in the r.h.s. of
(\ref{Result_LequalLp_plus})
and
(\ref{Result_LequalLp_minus})
goes only over even
$k$s.
Therefore,
$N-k$
is also even.
Now we work out the selection rule for
$J$:
since
$d^J_{\pm 1, 0}(\theta) \sim \sin \theta P'_J(\cos \theta)$,
only odd $J$s are consistent with the $T$-invariance.
From the covariance condition the highest possible value of 
$J$ 
for given
$N$
is
$J=N+1$.
There is no selection in parity $P$, since the operator in question does not possess definite parity
(see App.~\ref{Sec_Decomposition}).
So both $P=\pm 1$ exchanges are possible. Thus, {\it e.g.} for 
$N=0$ 
we recover
$J^{PC}=1^{--}$
and
$J^{PC}=1^{+-}$ exchanges quoted in second lines of
Tables~\ref{Table_NP_quarks}, 
\ref{Table_ANP_quarks} 
in Appendix~\ref{Sec_Decomposition}
while for
$N=2$
$J^{PC}=1^{--}, \, 3^{--}$
and
$J^{PC}=1^{+-}, \, 3^{+-}$
are relevant.

The combinations
(\ref{comb_without_definite})
can be considered according to same pattern.
However, to get the polynomials of definite parity properties in
$\cos \theta$
in order to work the selection rules in
$J$
one should consider the sum and difference of the two combinations in
(\ref{comb_without_definite}).
This analysis completes the set of
$J^{PC}$
exchanges
of Tables~\ref{Table_NP_quarks},
\ref{Table_ANP_quarks}  by including some of the contributions of unnatural parity meson
that arise only for combinations involving GPD $\tilde{E}_T^q$.
One can also check that for given $N$ the number of independent
generalized FFs (which is controlled by the power of the appropriate polynomials)
coincides with that from the last columns of
Tables~\ref{Table_NP_quarks},
\ref{Table_ANP_quarks}.

\subsection{${\rm SO}(3)$ partial wave decomposition of gluon helicity flip GPDs}
\label{SSec_dec_gluons}
In the complete analogy with our previous analysis we now proceed with the
${\rm SO}(3)$
partial wave decomposition of gluon helicity flip GPDs.
The  cross channel analytic continuation  of the form factor decomposition
(\ref{FF_dec_My})
of
$N$-th Mellin moments of gluon helicity flip GPDs is given by
\be
&&
   { \mathbb{S}}_{\{ \alpha \beta \mu_1...\mu_N\}}
   \langle N(p', \lambda') \bar{N}(\tilde{p}, \lambda)  |\hat{O}_{ T}^{g \, \alpha \rho \, \beta \sigma \, \mu_1  \ldots \mu_N}(0)| 0 \rangle
 \nonumber \\ &&
 = { \mathbb{ S}}_{\{ \alpha \beta \mu_1...\mu_N\}} \; \bar{U}(p', \lambda') \Big[ \sum_{k=0 \atop {\rm even}}^N
 \Big\{  \frac{ \frac{1}{2} \tilde{\Delta}^\beta \tilde{P}^\sigma- \frac{1}{2} \tilde{\Delta}^\sigma \tilde{P}^\beta}{2m}  i \sigma^{\alpha \rho}
\tilde{P}^{\mu_1} \ldots \tilde{P}^{\mu_k} \frac{1}{2} \tilde{\Delta}^{\mu_{k+1}} \ldots \frac{1}{2} \tilde{\Delta}^{\mu_{N}} A_{T \, N+1,k}^g(\tilde{s})  \nonumber \\ &&
 +\frac{\frac{1}{2} \tilde{\Delta}^\beta \tilde{P}^\sigma- \frac{1}{2} \tilde{\Delta}^\sigma \tilde{P}^\beta}{2m} \frac{\frac{1}{2} \tilde{\Delta}^\alpha \tilde{P}^\rho- \frac{1}{2} \tilde{\Delta}^\rho \tilde{P}^\alpha}{m^2}  \tilde{P}^{\mu_1} \ldots \tilde{P}^{\mu_k} \frac{1}{2} \tilde{\Delta}^{\mu_{k+1}} \ldots \frac{1}{2} \tilde{\Delta}^{\mu_{N}} \tilde{A}_{T \, N+1,k}^g(\tilde{s})
  \nonumber \\ &&
 + \frac{\frac{1}{2} \tilde{\Delta}^\beta \tilde{P}^\sigma- \frac{1}{2} \tilde{\Delta}^\sigma \tilde{P}^\beta}{2m}  \frac{\gamma^\alpha \tilde{P}^\rho- \gamma^\rho \tilde{P}^\alpha}{2m} \tilde{P}^{\mu_1} \ldots \tilde{P}^{\mu_k} \frac{1}{2} \tilde{\Delta}^{\mu_{k+1}} \ldots \frac{1}{2} \tilde{\Delta}^{\mu_{N}}  B_{T \, N+1,k}^g( \tilde{s}) \Big\}
   \nonumber \\ &&
+ \sum_{k=0 \atop {\rm odd}}^N
\frac{\frac{1}{2} \tilde{\Delta}^\beta \tilde{P}^\sigma- \frac{1}{2} \tilde{\Delta}^\sigma \tilde{P}^\beta}{2m}
 \frac{\gamma^\alpha \frac{1}{2} \tilde{\Delta}^\rho- \gamma^\rho \frac{1}{2} \tilde{\Delta} ^\alpha}{m^2} \tilde{P}^{\mu_1} \ldots \tilde{P}^{\mu_k} \frac{1}{2}
 \tilde{\Delta}^{\mu_{k+1}} \ldots \frac{1}{2} \tilde{\Delta}^{\mu_{N}}  \tilde{B}_{T \, N+1,k}^g( \tilde{s})
 \Big] V(\tilde{p},\lambda).
   \nonumber \\ &&
 \label{FF_dec_My_Cross}
\ee

We would like to compute the matrix elements of $N$-th Mellin moment of
gluon helicity flip
$N \bar{N}$
GDA
\be
\langle  N(p',\lambda') \bar{N}( \tilde{p},\lambda)| \hat{O}_{  T}^{g\, + i, +j , \, ++ \ldots + }|0 \rangle
\ee
in $N \bar{N}$ CMS.
We find the following combinations  with definite operator helicities $J_3$
suitable for PW expansion in the $t$-channel PW:
\be
&&
\langle  N(p',\lambda') \bar{N}( \tilde{p},\lambda)| \hat{O}_{  T}^{g\, + 1, +1 , \, ++ \ldots + }|0 \rangle-
\langle  N(p',\lambda') \bar{N}( \tilde{p},\lambda)| \hat{O}_{  T}^{g \, + 2, +2 , \, ++ \ldots + }|0 \rangle
\nonumber  \\ && + 2 i
\langle  N(p',\lambda') \bar{N}( \tilde{p},\lambda)| \hat{O}_{  T}^{g\,+ 1, +2 , \, ++ \ldots + }|0 \rangle
\nonumber  \\ &&
\equiv \langle  N(p',\lambda') \bar{N}( \tilde{p},\lambda)| \hat{O}_{  T}^{g\, + (1+i2), +(1+i2) , \, ++ \ldots + }|0 \rangle;
\ee
\be
&&
\langle  N(p',\lambda') \bar{N}( \tilde{p},\lambda)| \hat{O}_{  T}^{g\, + 1, +1 , \, ++ \ldots + }|0 \rangle-
\langle  N(p',\lambda') \bar{N}( \tilde{p},\lambda)| \hat{O}_{  T}^{g\,+ 2, +2 , \, ++ \ldots + }|0 \rangle
\nonumber  \\ && - 2 i
\langle  N(p',\lambda') \bar{N}( \tilde{p},\lambda)| \hat{O}_{  T}^{g\,+ 1, +2 , \, ++ \ldots + }|0 \rangle
\nonumber  \\ &&
\equiv \langle  N(p',\lambda') \bar{N}( \tilde{p},\lambda)| \hat{O}_{  T}^{g\,+ (1-i2), +(1-i2) , \, ++ \ldots + }|0 \rangle;
\ee
and
\be
&&
\langle  N(p',\lambda') \bar{N}( \tilde{p},\lambda)| \hat{O}_{  T}^{g\, + 1, +1 , \, ++ \ldots + }|0 \rangle+
\langle  N(p',\lambda') \bar{N}( \tilde{p},\lambda)| \hat{O}_{  T}^{g \, + 2, +2 , \, ++ \ldots + }|0 \rangle
\nonumber  \\ &&
\equiv \langle  N(p',\lambda') \bar{N}( \tilde{p},\lambda)| \hat{O}_{  T}^{g\, + (1+i2), +(1-i2) , \, ++ \ldots + }|0 \rangle.
\label{combinations_llp0}
 \ee
These combinations correspond to the operator helicities
$J_3=+2$,
$J_3=-2$
and
$J_3=0$
respectively.

Then for
$\lambda= \lambda'$
({\it i.e.} the aligned configuration of nucleon and antinucleon helicities:
$N^\uparrow \bar{N}^\uparrow$ or $N^\downarrow \bar{N}^\downarrow$)
we get
\be
&&
\left. \langle  N(p',\lambda') \bar{N}( \tilde{p},\lambda)| \hat{O}_{  T}^{g\, + (1+i2), +(1+i2) , \,  ++ \ldots +  }|0 \rangle \right|_{\lambda= \lambda'}
\nonumber  \\ &&
= \eta_{\lambda \lambda'}^{++} (\tilde{P}^+)^{N+2}  \frac{\beta}{2 \sqrt{1-\beta}}\sin^2 \theta \sum_{k=0 \atop \rm even}^N \Big[
(1-\beta) A_{T \, N+1,k}^g(\tilde{s})
\nonumber  \\ &&
+2 \left(1-\frac{\tilde{s}}{4m^2} \right) \tilde{A}_{T \, N+1,k}^g(s)+B_{T \, N+1,k}^g(\tilde{s})
\Big] \left( \frac{1}{2} \beta \cos \theta \right)^{N-k};
\ee
\be
&&
\left. \langle  N(p',\lambda') \bar{N}( \tilde{p},\lambda)| \hat{O}_{  T}^{g \, + (1-i2), +(1-i2) , \,  ++ \ldots +  }|0 \rangle\right|_{\lambda= \lambda'}
\nonumber  \\ &&
= \eta_{\lambda \lambda'}^{--} (\tilde{P}^+)^{N+2}  \frac{\beta}{2 \sqrt{1-\beta}}\sin^2 \theta \sum_{k=0 \atop \rm even}^N \Big[
(1+\beta) A_{T \, N+1,k}^g(\tilde{s})
\nonumber  \\ &&
+2 \left(1-\frac{\tilde{s}}{4m^2} \right) \tilde{A}_{T \, N+1,k}^g(\tilde{s})+B_{T \, N+1,k}^g(\tilde{s})
\Big] \left( \frac{1}{2} \beta \cos \theta \right)^{N-k}.
\ee
The $\theta$ dependence of the appropriate partial waves is given by
\be
d^J_{\pm 2, 0}(\theta)=  \frac{1}{\sqrt{(J-1)J(J+1)(J+2)} } \sin^2 \theta \, P''_J(\cos \theta).
\ee

Performing crossing to the direct channel (see discussion in
Sec.~\ref{SubSec_PWquark})
we  conclude that the combinations
\be
&&
(1-\tau) \tilde{H}_T^g +\frac{1}{2} E_T^g; \nonumber \\ &&
H_T^g + (1-\tau) \tilde{H}_T^g + \frac{1}{2} E_T^g
\label{comb23}
\ee
are to be expanded in
$ P''_J(1/\xi)$.

For the  combination
(\ref{combinations_llp0})
we  find
\be
&&
\left. \langle  N(p',\lambda') \bar{N}( \tilde{p},\lambda)| \hat{O}_{  T}^{g\, + (1+i2), +(1-i2) , \,  ++ \ldots +  }|0 \rangle \right|_{\lambda= \lambda'}
\nonumber \\ &&
=\eta_{\lambda \lambda'}^{+-} (\tilde{P}^+)^{N+2}  \frac{\beta}{4 \sqrt{1-\beta}}\sin^2 \theta \sum_{k=0 \atop \rm even}^N \Big[
\beta A_{T \, N+1,k}^g(\tilde{s})
\nonumber  \\ &&
+2 \left(1-\frac{\tilde{s}}{4m^2} \right) \tilde{A}_{T \, N+1,k}^g(\tilde{s})+B_{T \, N+1,k}^g(\tilde{s})
\Big] \left( \frac{1}{2} \beta \cos \theta \right)^{N-k}.
\ee
It is to be expanded in
\be
d^J_{0, 0}(\theta)=    P_J(\cos \theta).
\ee
Performing crossing to the direct channel we conclude that the combination
\be
(\xi^2 -1) \left( H_T^g +2(1-\tau) \tilde{H}_T^g +E_T^g \right)
\label{comb1}
\ee
is to be expanded in
$\xi^2 P_J(1/\xi)$.

Now we repeat the analysis for
$|\lambda'-\lambda|=1$
({\it i.e.} the opposite  helicity configuration of nucleon and antinucleon:
$N^\uparrow \bar{N}^\downarrow$ or $N^\downarrow \bar{N}^\uparrow$).
We find out that
\be
&&
\left. \langle  N(p',\lambda') \bar{N}(\tilde{p},\lambda)| \hat{O}_{  T}^{g\, + (1+i2), +(1+i2) , \,  ++ \ldots +  }|0 \rangle \right|_{|\lambda'-\lambda|=1}
\nonumber  \\ &&
= {\eta'}_{\lambda \lambda'}^{++} (\tilde{P}^+)^{N+2}  \frac{1}{2} \beta \sin  \theta (1+\cos \theta) \Big\{ \sum_{k=0 \atop \rm even}^N \Big[
 A_{T \, N+1,i}^g(\tilde{s})
\nonumber  \\ &&
+ \frac{\tilde{s}}{4m^2} B_{T \, N+1,i}^g(\tilde{s}) \Big] \left( \frac{1}{2} \beta \cos \theta \right)^{N-k} -
\sum_{k=0 \atop \rm odd}^N \beta \frac{\tilde{s}}{4m^2} \tilde{B}_{T \, N+1,k}^g(\tilde{s})
\left( \frac{1}{2} \beta \cos \theta \right)^{N-k} \Big\};
\ee
\be
&&
\left. \langle  N(p',\lambda') \bar{N}( \tilde{p},\lambda)| \hat{O}_{ T}^{g\, + (1-i2), +(1-i2) , \,  ++ \ldots +  }|0 \rangle  \right|_{|\lambda'-\lambda|=1}
\nonumber  \\ &&
= {\eta'}_{\lambda \lambda'}^{--} (\tilde{P}^+)^{N+2}  \frac{1}{2} \beta \sin  \theta ( \cos \theta-1) \Big\{ \sum_{k=0 \atop \rm even}^N \Big[
 A_{T \, N+1,k}^g(\tilde{s})
\nonumber  \\ &&
+ \frac{\tilde{s}}{4m^2} B_{T \, N+1,k}^g(\tilde{s}) \Big] \left( \frac{1}{2} \beta \cos \theta \right)^{N-k} +   \sum_{k=0 \atop \rm odd}^N \beta \frac{\tilde{s}}{4m^2} \tilde{B}_{T \, N+1,k}^g(\tilde{s})
\left( \frac{1}{2} \beta \cos \theta \right)^{N-k} \Big\}
\ee
are to be expanded respectively in
\be
&&
d^J_{2, 1}(\theta)
\nonumber \\ &&
=\sqrt{\frac{1}{(J-1) (J+2)}} \frac{1 }{J (J+1)} \sin \theta  (\cos \theta +1) \left( 2 P''_J(\cos \theta) +(\cos \theta-1) P'''_J(\cos \theta)  \right)
\nonumber  \\ &&
\ee
and
\be
&&
d^J_{- 2, 1}(\theta)
\nonumber \\ &&
=\sqrt{\frac{1}{(J-1) (J+2)}} \frac{1 }{J (J+1)} \sin \theta  (\cos \theta -1) \left( 2 P''_J(\cos \theta) +(\cos \theta+1) P'''_J(\cos \theta)  \right)
\nonumber  \\ &&
\ee

Crossing back to the direct channel
we conclude that the combinations
\be
H_T^g+ \tau E_T^g  \pm   \tau \xi \tilde{E}_T^g
\ee
are to be expanded in
$2 P''_J(1/\xi) +\frac{1 \mp \xi}{\xi} P'''_J(1/\xi)$.

Finally, the combination
\be
&&
\left. \langle  N(p',\lambda') \bar{N}( \tilde{p},\lambda)| \hat{O}_{g T}^{+ (1+i2), +(1-i2) , \,  ++ \ldots +  }|0 \rangle  \right|_{|\lambda'-\lambda|=1}
\nonumber  \\ &&
= {\eta'}_{\lambda \lambda'}^{+-} (\tilde{P}^+)^{N+2}  \frac{1}{4}  \sin  \theta \Big\{ \beta \cos \theta  \sum_{k=0 \atop \rm even}^N \Big[
 A_{T \, N+1,k}^g(\tilde{s})
+ \frac{\tilde{s}}{4m^2} B_{T \, N+1,k}^g(\tilde{s}) \Big] \left( \frac{1}{2} \beta \cos \theta \right)^{N-k} \nonumber  \\ &&
- \beta^2  \sum_{k=0 \atop \rm odd}^N \frac{\tilde{s}}{4m^2} \tilde{B}_{T \, N+1,i}^g(\tilde{s})
\left( \frac{1}{2} \beta \cos \theta \right)^{N-k}
\Big\}
\nonumber  \\ &&
\ee
is to be expanded in
\be
d^J_{ 0, 1}(\theta) =  \frac{1}{\sqrt{ J(J+1)} } \sin \theta \, P'_J(\cos \theta).
\ee
After crossing back to the direct channel
it has to do with the combination
\be
H_T^g+ \tau E_T^g +   \tau \xi \tilde{E}_T^g
\label{comb23bis}
\ee
that is to be expanded in $P'_J(1/\xi)$.

Thus, we found out the combinations of gluon helicity flip GPDs suitable for the PW expansion in the 
cross-channel partial waves.
Moreover, similarly to the quark case, from comparing the
explicit expressions for the Mellin moments of gluon helicity GDA
one may work out the
$J^{PC}$
selection rules for $t$-channel exchanges contributing to the Mellin moments of
gluon helicity flip GPDs. The results are summarized in
Tables~\ref{Table_NP_gluons}, \ref{Table_ANP_gluons} 
of Appendix~\ref{Sec_Decomposition}.
One may check  these selection rules
are consistent with those worked out by the method 
\cite{Ji:2000id}
(see Sec.~\ref{SubSec_QN_gluons}).
Let us emphasize that, due to the fact that gluon is its own antiparticle, gluon helicity flip GPDs
are all $C$-even. In other words, gluon helicity flip GPDs are even functions of the
variable
$x$ 
\cite{Diehl:2001pm}
\footnote{Therefore, we were a bit puzzled by the consideration of odd Mellin moments of
gluon helicity flip GPDs in
Ref.~\cite{Chen:2004cg},
that all should be zero.}.
In our Tables~\ref{Table_NP_gluons}, \ref{Table_ANP_gluons}
we show the results only for even Mellin moments of gluon helicity flip GPDs.

\section{On the cross channel meson exchange contributions  to quark and gluon helicity flip GPDs}
\label{Sec_Resonance}

The alternative method to establish the set of quark and gluon helicity flip GPDs suitable for the
partial wave expansion in the cross-channel partial waves consists in the explicit calculation of
the cross channel spin-$J$ resonance contributions into corresponding GPD
(see \cite{Polyakov:1999gs}).
The advantage of this method is that it is fully covariant and allows to determine the net
resonance exchange contributions into scalar invariant functions
$H_{T}^{q,g}$,
$E_{T}^{q,g}$,
$\tilde{H}_{T}^{q,g}$,
$\tilde{E}_{T}^{q,g}$.
In particular, the transition to the limit of massless hadrons can then be performed for
the partial wave expansion of those scalar functions
(and not for the operator matrix elements as in the analysis of
Sec.~\ref{Sec_SO(3)exp}).
This method also allows to work out the explicit expressions for the double partial wave expansion of
quark and gluon helicity flip GPDs within the dual parametrization approach. Therefore, we find it useful
to present below a short overview of this approach.

\begin{figure}[H]
 \begin{center}
 \epsfig{figure=  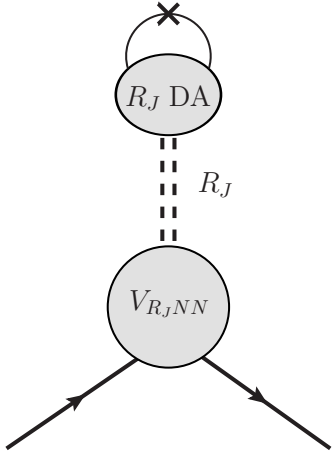 , height=6cm}
      \caption{
      Spin-$J$ resonance exchange contribution into the nucleon matrix element of light-cone operator
      $\hat{O}$.
      The upper blob denotes the appropriate distribution amplitude of spin-$J$ meson resonance. The lower blob
      denotes the minimal (on-shell) effective $R_J N N$ vertex.
      }
\label{Fig_Rex}
\end{center}
\end{figure}

The calculation is based on the assumption of the meson pole dominance for the matrix element of the light-cone operator in question.
This can be seen as the generalization of the old idea of vector meson dominance for nucleon electromagnetic form factor.
In principle, this picture can be justified in the large-$N_c$ limit of QCD in which meson resonances become infinitely narrow.
What is more important for us is that, by assuming the most general form of nucleon-meson interaction consistent with a given set
of symmetries, we populate the invariant form factors at all possible leading twist spin-tensor structures and recover the usual set
of selection rules for the quantum numbers of the cross channel resonance exchanges. Within this approach
the nucleon matrix element of the light-cone operator
$\hat{O}$
is presented as the infinite sum of contributions depicted on Fig.~\ref{Fig_Rex}.
Symbolically it can be written in the following form:
\be
&&
\langle N(p')|\,
\hat{O} \,
| N(p) \rangle
\nonumber \\ &&
\sim
\sum_{R_J}
\sum_{{\rm polarizations} \atop {\rm of \;} R_J }
\frac{1}{\Delta^2-M_{R_J}^2}
\times
\underbrace{\; \langle N(p') R_J(\Delta) | N(p)  \rangle \; }_{V_{R_J N  N}  {\rm \; effective \; vertex} } \ \ \ \otimes
\underbrace{ \langle 0 | \,
\hat{O}  \,
| \, R_J(\Delta)\rangle }_{ {\rm Fourier \; Transform\; of \; DA \; of \;} R_J},
\label{Ma trix_element}
\ee
where
$M_{R_J}$ stand for the resonance masses and
$\otimes$ denotes the convolution in the appropriate Lorentz indices.
Since we are interested just in the pole contributions to the matrix elements it
suffices to use the so-called minimal forms (those which do not change their form
when all particles are put on mass shell) of
$R_J N N$ effective vertices (see
Ref.~\cite{SemenovTianShansky:2005ky,SemenovTianShansky:2007hv,Semenov-Tian-Shansky:2013qva}
for the detailed discussion).
The resulting on-shell polarization sums for spin-$J$ resonances can be performed with the contracted projectors method
(see {\it e.g.} Chapter I of 
\cite{Alfaro_red_book}).

Now we discuss the application of the approach for the case of quark and gluon helicity flip GPDs, which represents the new result.
The effective minimal
$R_J \bar{N} N$
vertices for natural parity 
($P=(-1)^J$) 
mesons can be chosen in the following form
\cite{SemenovTianShansky:2007hv}:
\be
V_{R_J N N}=
\bar{U}(p')
\Big\{ \frac{g_1^{R_J}}{M_{R_J}^{J-1}}   \gamma^{\mu_1 } P^{\mu_2}...P^{\mu_J}
+ \frac{g_2^{R_J}}{M_{R_J}^J}  P^{\mu_1}...P^{\mu_J}
\Big\} U(p) \, {\cal E}^*_{\mu_1 ... \mu_J}(\Delta,j),
\label{effvert_NormalP}
\ee
where
$g_{1,2}^{R_J}$ are  dimensionless coupling constants and
${\cal E}_{\mu_1 ... \mu_J}(\Delta,j)$
stand for the symmetric traceless tensors describing spin-$J$ resonance polarization vector.

To keep with the requirements of $T$-invariance the effective minimal vertices for
$\tilde{R}_J^- \bar{N} N$
for the unnatural parity
($P=(-1)^{J+1}$)
mesons  with
$P \cdot C=-1$
({\it i.e.} $J^{PC}=1^{+-},\,2^{-+}, \, 3^{+-}, \ldots $  )
are chosen in a way that
\be
V_{\tilde{R}_J^- \bar{N} N}=   \frac{g^{\tilde{R}_J^-}}{M^{J}}
\bar{U}(p')  i \sigma_{\mu \nu} \gamma_5 U(p) P^{\mu_1}...P^{\mu_{J-1}} \Delta^\nu \, {\cal E}^*_{\mu \mu_1 ... \mu_{J-1}}(\Delta,j).
\label{VertexUnnatPTminus}
\ee
where
$g^{\tilde{R}_J^-}$
are   dimensionless.

Finally, the effective minimal vertices for
$\tilde{R}_J^+ \bar{N} N$
for the unnatural parity
($P=(-1)^{J+1}$)
mesons  with
$P \cdot C= 1$
 (i.e. with  $J^{PC}=1^{++},\,2^{--}, \, 3^{++}, \ldots $)
are chosen as
\be
V_{\tilde{R}_J^+ \bar{N} N}=   \frac{g^{\tilde{R}_J^+}}{M^{J}}
\bar{U}(p') i \sigma_{\mu \nu} \gamma_5 U(p) P^{\mu_1}...P^{\mu_{J-1}} P^\nu {\cal E}^*_{\mu \mu_1 ... \mu_{J-1}}(\Delta,j).
\label{VertexUnnatPTplus}
\ee

Other necessary ingredients for the calculation of
(\ref{Ma trix_element})
are the relevant leading twist DAs of mesons.
The corresponding parametrization for arbitrary high $J$
can be constructed analogously to vector meson (see {\it e.g.} 
\cite{Ball:1998sk})
and tensor meson
\cite{Braun:2000cs}
cases.

Namely, the quark helicity flip DA for the natural parity mesons
$J^{PC}=1^{-\pm},\,2^{+\pm},\,3^{-\pm},\,4^{+\pm}$
will read as
\be
&&
\langle 0|
\bar{\Psi} (-  \lambda n/2) i \sigma^{+ i}   \Psi(  \lambda n/2)\, |R_J(\Delta,j) \rangle=
n^\alpha {g_\bot}^{i \beta} \otimes
  f^q_{T \, R_J}
 \frac{M^{J-1}}{(\Delta \cdot n)^{J-1}}
 \nonumber \\ &&
 \times
 n^{\nu_1}....n^{\nu_{J-1}}
 \left(
 \mathcal{E}_{\alpha \nu_1\,... \nu_{J-1}}(\Delta,j) \, \Delta_\beta-
 \mathcal{E}_{\beta \nu_1\,... \nu_{J-1}}(\Delta,j)\, \Delta_\alpha
 \right)
 \int_{-1}^1 dy e^{i y \frac{\lambda \Delta \cdot  n}{2}} \Phi^q_{T \, R_J} (y).
 \nonumber \\ &&
 \label{DAqNP}
\ee
The distribution amplitudes
$\Phi^q_{T \, R_J}(y)$
can be expanded over the basis of Gegenbauer polynomials
$C_n^{3/2}(y)$
as
\be
\Phi^q_{T \, R_J} (y )= (1-y^2) \frac{2^J \Gamma(J+\frac{1}{2})}{\Gamma{(\frac{1}{2})} \Gamma{(J+2)}}
\sum_{k=J-1 \atop \rm odd/\,even }^\infty (a^q_{T \, R_J})_k  \, C_k^{\frac{3}{2}} (y) \ \ \text{with} \ \ (a^q_{T \, R_J})_{J-1}=1.
\label{DA_Normal_P}
\ee
The expansion
(\ref{DA_Normal_P})
runs over even
$k$
for
$C=-1$
mesons and over odd
$k$
for
$C=+1$
mesons. The normalization of the DA
(\ref{DA_Normal_P})
is chosen such that
\be
\int_{-1}^1 dy y^{J-1}\Phi^q_{T \, R_J} (y)=1\,.
\label{norm_DA}
\ee
The normalization constant
$ f^q_{T \, R_J}$
has the dimension of energy.

For the case of the unnatural parity mesons
$J^{PC}=1^{+\pm},
\,2^{-\pm},\,
3^{+\pm},
\,4^{-\pm},
...$
the 
quark helicity flip DAs take the form
\be
&&
\langle 0|
\bar{\Psi} (-\lambda n/2) \sigma^{+ i} \gamma_5 \Psi(\lambda n/2)\, |\tilde{R}_J(\Delta,j) \rangle=
n^\alpha {g_\bot}^{i \beta} \otimes i  f^q_{T \, \tilde{R}_J}
 \frac{M^{J-1}}{(\Delta \cdot n)^{J-1}}
 \nonumber \\ &&
 \times n^{\nu_1}....n^{\nu_{J-1}}
 \left(
 \mathcal{E}_{\alpha \nu_1\,... \nu_{J-1}}(\Delta,j) \, \Delta_\beta-
 \mathcal{E}_{\beta \nu_1\,... \nu_{J-1}}(\Delta,j)\, \Delta_\alpha
 \right)
 \int_{-1}^1 dy e^{i y \frac{\lambda \Delta \cdot n}{2}} \Phi^q_{T \, \tilde{R}_J}(y )\,.
 \nonumber \\ &&
 \label{DA_uunatP}
\ee
The Gegenbauer expansion for
$ \Phi^q_{T \, \tilde{R}_J}(y )$  is the same as
(\ref{DA_Normal_P}).
Again  the sum runs over even $k$ for $C=-1$
mesons and over odd
$k$
for
$C=+1$
mesons. The normalization of
$ \Phi^q_{T \, \tilde{R}_J}(y )$
is the same as
(\ref{norm_DA})
with
$ f^q_{T \, \tilde{R}_J}$
also having the dimension of energy.

For the case of the gluon helicity flip operator the relevant DAs for
the natural parity even spin mesons 
($J^{PC}=2^{++}, 4^{++}, \ldots$) 
read
\be
&&
\langle 0|  {\mathbb{S}} G^{+ i} (-\lambda n/2 ) G^{j + } (\lambda n/2) |  R_J(\Delta,j) \rangle
\nonumber \\ &&
= n^\alpha n^\beta \tau^\bot_{ij; \, \rho \sigma} \otimes  \;
{f^g_{T \, R_J}} \,  M_R^{J-2}   \left\{ \frac{1}{4} \Delta^\alpha \Delta^\beta {\cal E}^{\rho \sigma  \nu_1...\nu_{J-2} }(\Delta,j)  -
 \frac{1}{4} \Delta^\rho \Delta^\beta {\cal E}^{\alpha \sigma  \nu_1...\nu_{J-2}}(\Delta,j) \right. \nonumber \\ && \left.
 - \frac{1}{4} \Delta^\alpha \Delta^\sigma {\cal E}^{\rho \beta  \nu_1...\nu_{J-2} }(\Delta,j) +
 \frac{1}{4} \Delta^\rho \Delta^\sigma {\cal E}^{\alpha \beta \nu_1...\nu_{J-2}}(\Delta,j)  \right\}
  n^{\nu_1}....n^{\nu_{J-2}} \left(  \frac{2}{(\Delta \cdot n)} \right)^{J-2}
  \nonumber \\ &&
  \times
 \int_{-1}^1 dy e^{i y \lambda \frac{\Delta \cdot n}{2}} \Phi^g_{T \, R_J}(y),
\label{DA_high_J}
\ee
where
the DA
$\Phi^g_{T \, R_J}(y)$
is in general expanded over the set of Gegenbauer polynomials
$C_k^{5/2}(y)$:
\be
\Phi^g_{T \, R_J}(y)= \frac{3\ 2^J \Gamma \left(J+\frac{3}{2}\right)}{\Gamma(\frac{1}{2})
 \Gamma (J+3)} (1-y^2)^2   \sum_{k=J-2 \atop {\rm even}}^\infty (a^g_{T \, R_J})_k C_k^{\frac{5}{2}}(y)
\ \ {\rm with} \ \
(a^g_{T \, R_J})_{J-2}=1.
\label{Sum_geg_gluons}
\ee
It is normalized so that
\be
\int_{-1}^1 dy y^{J-2} \Phi^g_{T \, R_J}(y)= 1
\ee
and the normalization constant
$f^g_{T \, R_J}$
has the dimension of energy.
Note, that the sum in
(\ref{Sum_geg_gluons})
runs over even
$k$
since we deal with the
$C$-even
quantity, because the gluon is its own antiparticle.

Finally, for unnatural parity mesons 
($J^{PC}=2^{-+}, \, 3^{++}, \, 4^{-+} \ldots$)  
the relevant DAs read ad
\be
&&
\langle 0|  {\mathbb{S}} \tilde{G}^{+ i} (-\lambda n/2 ) G^{j + } (\lambda n/2) |  R_J(\Delta,j) \rangle
\nonumber \\ &&
= n^\alpha n^\beta \tau^\bot_{ij; \, \rho \sigma} \otimes  \;
i {f^g_{T \, \tilde{R}_J}} \,  M_R^{J-2}   \left\{ \frac{1}{4} \Delta^\alpha \Delta^\beta {\cal E}^{\rho \sigma  \nu_1...\nu_{J-2} }(\Delta,j)  -
 \frac{1}{4} \Delta^\rho \Delta^\beta {\cal E}^{\alpha \sigma  \nu_1...\nu_{J-2}}(\Delta,j) \right. \nonumber \\ && \left.
 - \frac{1}{4} \Delta^\alpha \Delta^\sigma {\cal E}^{\rho \beta  \nu_1...\nu_{J-2} }(\Delta,j) +
 \frac{1}{4} \Delta^\rho \Delta^\sigma {\cal E}^{\alpha \beta \nu_1...\nu_{J-2}}(\Delta,j)  \right\}
  n^{\nu_1}....n^{\nu_{J-2}} \left(  \frac{2}{(\Delta \cdot n)} \right)^{J-2}
  \nonumber \\ &&
  \times
 \int_{-1}^1 dy e^{i y \lambda \frac{\Delta \cdot n}{2}} \Phi^g_{T \, \tilde{R}_J}(y),
\label{DA_high_J_unnatP}
\ee
where for even 
$J$ 
the DA 
$\Phi^g_{T \, \tilde{R}_J}(y)$ 
has the same Gegenbauer expansion as in
(\ref{Sum_geg_gluons}).
For odd
$J$
the  Gegenbauer expansion reads
\be
\Phi^g_{T \, \tilde{R}_J}(y)= \frac{3\ 2^J \Gamma \left(J+\frac{3}{2}\right)}{\Gamma(\frac{1}{2}) \Gamma (J+3)} (1-y^2)^2   
\sum_{k=J-3 \atop {\rm even}}^\infty (a^g_{T \, \tilde{R}_J})_k C_k^{\frac{5}{2}}(y)
\ \ {\rm with} \ \
(a^g_{T \, \tilde{R}_J})_{J-3}=1.
\label{Sum_geg_gluons_oddJ_uunatP}
\ee
It is normalized so that
\be
\int_{-1}^1 dy y^{J-3} \Phi^g_{T \, \tilde{R}_J}(y)= 1.
\ee

Using the set of effective vertices
(\ref{effvert_NormalP}),
(\ref{VertexUnnatPTminus}),
(\ref{VertexUnnatPTplus})
and the
parton helicity flip DAs
(\ref{DAqNP}),
(\ref{DA_uunatP}),
(\ref{DA_high_J}),
(\ref{DA_high_J_unnatP})
after a straightforward (though tedious and lengthy)
calculation one recovers the same combinations of quark and gluon helicity flip
GPDs suitable for the partial wave expansion in the cross channel partial waves as
derived in Sec.~\ref{Sec_SO(3)exp}
and may check the selection rules for the
$J^{PC}$
quantum numbers.

Moreover, one can work out the double partial wave expansion (both in conformal partial waves and in cross channel
${\rm SO(3)}$
partial waves%
\footnote{Below we relabel the cross channel angular momentum as $l$ to match the notations of 
Ref.~\cite{Polyakov:2002wz}.}.)
representing quark and gluon helicity flip GPDs within the dual parametrization approach. As an example
we present below the formal series for the non-singlet
($C=-1$)
combinations of quark helicity flip GPDs within the dual parametrization approach. For simplicity we quote the results
in the 
$\beta \to 1$ 
limit.
\be
&&
\tau
\tilde{H}_T^{q^-}(x,\xi,\Delta^2) -\frac{1}{2} E_T^{q^-}(x,\xi,\Delta^2)
\nonumber \\ &&
= \sum_{k=0 \atop {\rm even }}^\infty \sum_{l=1 \atop {\rm odd}}^{k+1}
B_{kl}^{(\tau
\tilde{H}_T^{q^-}  -\frac{1}{2} E_T^{q^-})} (\Delta^2)
\theta \left(1- \frac{x^2}{\xi^2} \right)  \left(1- \frac{x^2}{\xi^2} \right) C_k^{3/2} \left( \frac{x}{\xi} \right) \frac{1}{|\xi|} P'_l \left( \frac{1}{\xi} \right)
\nonumber \\ &&
-  H_T^{q^-}(x,\xi,\Delta^2)+
\tau
\tilde{H}_T^{q^-}(x,\xi,\Delta^2) -\frac{1}{2} E_T^{q^-}(x,\xi,\Delta^2)
\nonumber \\ &&
= \sum_{k=0 \atop {\rm even }}^\infty \sum_{l=1 \atop {\rm odd}}^{k+1}
B_{kl}^{(-  H_T^{q^-}+
\tau
\tilde{H}_T^{q^-} -\frac{1}{2} E_T^{q^-})} (\Delta^2)
\theta \left(1- \frac{x^2}{\xi^2} \right)  \left(1- \frac{x^2}{\xi^2} \right) C_k^{3/2} \left( \frac{x}{\xi} \right) \frac{1}{|\xi|} P'_l \left( \frac{1}{\xi} \right)
\nonumber \\ &&
H_T^{q^-}(x,\xi,\Delta^2)+ 
\tau \tilde{H}_T^{q^-}(x,\xi,\Delta^2)  \pm
\tau \tilde{E}_T^{q^-}(x,\xi,\Delta^2)
\nonumber \\ &&
=\sum_{k=0 \atop {\rm even }}^\infty \sum_{l=1 }^{k+1}
B_{kl}^{(H_T^{q^-} +
\tau \tilde{H}_T^{q^-}  \pm
\tau \tilde{E}_T^{q^-} )} (\Delta^2)
\theta \left(1- \frac{x^2}{\xi^2} \right)  \left(1- \frac{x^2}{\xi^2} \right) C_k^{3/2} \left( \frac{x}{\xi} \right) \frac{1}{|\xi|}
\nonumber \\ && \times
\left( P'_l \left( \frac{1}{\xi} \right)   +   \frac{1 \mp \xi}{\xi}  P''_l \left( \frac{1}{\xi} \right) \right).
\label{Dual_par_quarks}
\ee
Here we introduce
$4$
sets of generalized form factors
$B_{kl}(\Delta^2)$
for the $4$ combinations in question. Note, that the sum in $l$ for the combinations
$H_T^{q^-} +
\tau \tilde{H}_T^{q^-}  \pm
\tau \tilde{E}_T^{q^-}$
runs over both odd and even $l$. This is consistent with the selection rules summarized in Table~{\ref{Table_ANP_quarks}}.
As a result both even and odd power of
$\xi$
will appear when computing even Mellin moments of these combinations
from the partial wave expansion
(\ref{Dual_par_quarks}).
This is obviously consistent with $T$-invariance since these
combinations explicitly contain the GPD
$\tilde{E}_T^{q^-}$
that produces odd powers of
$\xi$
for the Mellin moments (see Appendix~\ref{App_A}).

The formal series
(\ref{Dual_par_quarks})
can be handled by means of standard methods developed
within the dual parametrization approach
\cite{Polyakov:2002wz,Polyakov:2008aa,SemenovTianShansky:2010zv}.
The detailed consideration of these series however
lies beyond the scope of this paper and will be considered elsewhere.
The gluon case can be considered exactly the same pattern. However, the resulting expressions 
are bulky and we do not present them explicitly in the present publication.

\section{Conclusions}
\label{Sec_Conclusions}

Much work remains to be done before the helicity flip sector of quark and gluon generalized parton
distributions in the nucleon is understood. In this work, we established their crossed channel properties, which are of
direct importance for a theoretically consistent model building  in the spirit of the double partial
wave expansion of GPDs. We did not address  the phenomenological side of this study. Some experimental hints already point to
the observability of transversity quark and gluon GPDs.
The
$ (3\phi)$
modulation of the interference contribution to the unpolarized beam-longitudinally polarized target asymmetry seen in the HERMES data
\cite{Airapetian:2010aa}
for deeply virtual Compton scattering on a nucleon call for a significant gluon transversity contribution.
The recent transverse target spin asymmetries measured by COMPASS in vector meson exclusive leptoproduction
\cite{Adolph:2013zaa}
have been interpreted
\cite{Goloskokov:2013mba}
as a signal for transversity quark contributions. We shall cope with the phenomenology of quark and gluon
transversity GPDs in a future publication.

\section*{Acknowledgements}
We are  grateful to A. Belitsky, V. Braun, M. Diehl, D. Mueller and M. Vanderhaeghen for very helpful discussions and correspondence.
This work is partly supported by the Polish Grant NCN
No DEC-2011/01/B/ST2/03915,  the Joint Research Activity "Study of Strongly
Interacting Matter" (acronym HadronPhysics3, Grant 283286) under the Seventh
Framework Programme of the European Community, by the COPIN-IN2P3 Agreement,
by the French grant ANR PARTONS (ANR-12-MONU-0008-01)
and by the Tournesol 2014 Wallonia-Brussels-France Cooperation Programme.

\setcounter{section}{0}
\setcounter{equation}{0}
\renewcommand{\thesection}{\Alph{section}}
\renewcommand{\theequation}{\thesection\arabic{equation}}

\section{Form factor decomposition of the nucleon matrix elements of helicity flip operators}
\label{App_A}
\subsection{Quark helicity flip operators}
The form factor decomposition of the $N$-th Mellin moment of non-local tensor operator
(\ref{TensorOp})
is given by eq. (22) of Ref.~\cite{Hagler:2004yt}:
\be
&&
 { \mathbb{S}}_{\{ \nu \mu_1...\mu_N\}} 
 \langle p' | \bar{\Psi}(0) i \sigma^{\mu \nu} (\overleftrightarrow{i D}_{\mu_1})...(\overleftrightarrow{i D}_{\mu_N}) \Psi(0)| p \rangle
 \nonumber \\ &&
 = {\mathbb{S}}_{\{ \nu \mu_1...\mu_N\}} \; \bar{U}(p') \Big[ \sum_{k=0 \atop {\rm even}}^N
 \Big\{ i \sigma^{\mu \nu} \Delta^{\mu_1} \ldots \Delta^{\mu_k} P^{\mu_{k+1}} \ldots P^{\mu_{N}} A_{T \, N+1,k}^q(\Delta^2)  \nonumber \\ &&
 +\frac{P^\mu \Delta^\nu- P^\nu \Delta^\mu}{m^2} \Delta^{\mu_1} \ldots \Delta^{\mu_k} P^{\mu_{k+1}} \ldots P^{\mu_{N}} \tilde{A}_{T \, N+1,k}^q(\Delta^2)
  \nonumber \\ &&
 +  \frac{\gamma^\mu \Delta^\nu- \gamma^\nu \Delta^\mu}{m^2} \Delta^{\mu_1} \ldots \Delta^{\mu_k} P^{\mu_{k+1}} \ldots P^{\mu_{N}} B_{T \, N+1,k}^q(\Delta^2) \Big\}
   \nonumber \\ &&
+ \sum_{k=0 \atop {\rm odd}}^N
 \frac{\gamma^\mu P^\nu- \gamma^\nu P^\mu}{m^2} \Delta^{\mu_1} \ldots \Delta^{\mu_k} P^{\mu_{k+1}} \ldots P^{\mu_{N}} \tilde{B}_{T \, N+1,k}^q(\Delta^2)
 \Big] U(p),
 \label{FF_dec_Haegler}
\ee
where
$ {\mathbb{S}  }_{\{ \nu \mu_1...\mu_N\}}$
stands for the symmetrization and subsequent trace subtraction in the corresponding Lorentz indices.
The polynomiality relation for quark helicity flip GPDs then reads:
\be
&&
\int_{-1}^1 dx x^N H_T^q(x, \xi ,\Delta^2)= \sum_{k=0 \atop {\rm even}}^N (-2 \xi)^k A^q_{T \, N+1,k}(\Delta^2);  \nonumber \\ &&
\int_{-1}^1 dx x^N E_T^q(x, \xi ,\Delta^2)= \sum_{k=0 \atop {\rm even}}^N (-2 \xi)^k B^q_{T \, N+1,k}(\Delta^2); \nonumber \\ &&
\int_{-1}^1 dx x^N\tilde{H}_T^q(x, \xi ,\Delta^2)= \sum_{k=0 \atop {\rm even}}^N (-2 \xi)^k \tilde{A}^q_{T \, N+1,k}(\Delta^2); \nonumber \\ &&
\int_{-1}^1 dx x^N \tilde{E}_T^q(x, \xi ,\Delta^2)= \sum_{k=0 \atop {\rm odd}}^N  (-2 \xi)^k \tilde{B}^q_{T \, N+1,k}(\Delta^2).
\ee
Note, that in accordance with the requirements of the $T$-invariance the $N=0$ Mellin moment of GPD
$\tilde{E}_T^q$
vanishes. This fact is consistent with the counting of number of independent form factors for the
$N=0$
Mellin moment of quark helicity flip GPDs
(see Tables~\ref{Table_NP_quarks}, \ref{Table_ANP_quarks}).

\subsection{Gluon  helicity flip operators}
The form factor decomposition of the
$N$-th ($N$-even)
Mellin moment of non-local gluon tensor operator
(\ref{op_free_index})
is given by
({\it c.f.} eq.~(10) of
Ref.~\cite{Chen:2004cg}):
\be
&&
 {  \mathbb{S}}_{\{ \alpha \beta \mu_1...\mu_N\}}
 \langle p' |
 G^{\alpha \rho}(0)
 (\overleftrightarrow{i D}_{\mu_1})...(\overleftrightarrow{i D}_{\mu_N})
 G^{\beta \sigma}( 0)
 | p \rangle
 \nonumber \\ &&
 = {  \mathbb{S}}_{\{ \alpha \beta \mu_1...\mu_N\}} \; \bar{U}(p') \Big[ \sum_{k=0 \atop {\rm even}}^N
 \Big\{  \frac{P^\beta \Delta^\sigma- P^\sigma \Delta^\beta}{2m}  i \sigma^{\alpha \rho}
 \Delta^{\mu_1} \ldots \Delta^{\mu_k} P^{\mu_{k+1}} \ldots P^{\mu_{N}} A_{T \, N+1,k}^g(\Delta^2)  \nonumber \\ &&
 +\frac{P^\beta \Delta^\sigma- P^\sigma \Delta^\beta}{2m} \frac{P^\alpha \Delta^\rho- P^\rho \Delta^\alpha}{m^2}  \Delta^{\mu_1} \ldots \Delta^{\mu_k} P^{\mu_{k+1}} \ldots P^{\mu_{N}} \tilde{A}_{T \, N+1,k}^g(\Delta^2)
  \nonumber \\ &&
 + \frac{P^\beta \Delta^\sigma- P^\sigma \Delta^\beta}{2m}  \frac{\gamma^\alpha \Delta^\rho- \gamma^\rho \Delta^\alpha}{2m} \Delta^{\mu_1} \ldots \Delta^{\mu_k} P^{\mu_{k+1}} \ldots P^{\mu_{N}} B_{T \, N+1,k}^g(\Delta^2) \Big\}
   \nonumber \\ &&
+ \sum_{k=0 \atop {\rm odd}}^N
\frac{P^\beta \Delta^\sigma- P^\sigma \Delta^\beta}{2m}
 \frac{\gamma^\alpha P^\rho- \gamma^\rho P^\alpha}{m^2} \Delta^{\mu_1} \ldots \Delta^{\mu_k} P^{\mu_{k+1}} \ldots P^{\mu_{N}} \tilde{B}_{T \, N+1,k}^g(\Delta^2)
 \Big] U(p).
 \label{FF_dec_My}
\ee
Applying the projector $\tau_{i j; \, \rho \sigma}^\bot$ and taking the $+$-components in the indices
$\alpha, \, \beta, \, \mu_1, \ldots \mu_N$
we get for even
$N$
\be
&&
\int_{-1}^1 dx x^N H_T^g(x,\xi,t)= \sum_{k=0 \atop {\rm even}}^N A_{T \, N+k,i}^g(\Delta^2) \xi^k; \nonumber \\ &&
\int_{-1}^1 dx x^N \tilde{H}_T^g(x,\xi,t)= \sum_{k=0 \atop {\rm even}}^N \tilde{A}_{T \, N+1,k}^g(\Delta^2) \xi^i; \nonumber \\ &&
\int_{-1}^1 dx x^N E_T^g(x,\xi,t)= \sum_{k=0 \atop {\rm even}}^N B_{T \, N+1,k}^g(\Delta^2) \xi^k ;\nonumber \\ &&
\int_{-1}^1 dx x^N \tilde{E}_T^g(x,\xi,t)= \sum_{k=0 \atop {\rm odd}}^N \tilde{B}_{T \, N+1,k}^g(\Delta^2) \xi^k .\nonumber \\ &&
\ee
Again in accordance with the requirements of the $T$-invariance the $N=0$ Mellin moment of GPD
$\tilde{E}_T^g$
vanishes.

\setcounter{equation}{0}

\section{Form factor decomposition of twist-$2$ quark and gluon helicity flip operators and quantum number selection rules}
\label{Sec_Decomposition}

Polynomiality property of the Mellin moments of hadronic matrix elements of bilocal twist-$2$ QCD operators
is the direct manifestation of the Lorentz symmetry of the underlying fundamental quantum field theory.
The Mellin moments of the hadronic matrix elements of bilocal twist-$2$ QCD operators give rise to towers
of local twist-$2$ operators. These towers of local operators are parameterized in terms of generalized form factors.
The parametrization in question should also properly take into account  the consequences of the discrete
$P$
and
$T$
symmetries, which may reduce the number of independent generalized form factors. Therefore, establishing the
correct number of  independent form factors with proper implementation of constrains following from the discrete
$P$
and
$T$
symmetries provides the crucial consistency check. Moreover, this kind of analysis
allows one to work out the set of selection rules for
$J^{PC}$
quantum numbers of the $t$-channel exchange
resonances contributing into the Mellin moments  of twist-$2$ quark and gluon helicity flip GPDs.

The extensive analysis of the tensorial properties and form factor decomposition
of the leading twist quark and gluon helicity flip operators was carried in
Refs.~\cite{Hagler:2004yt,Geyer:1999uq,Ji:2000id}
employing the general method elaborated in
\cite{Ji:2000id}.
The key idea of the method \cite{Ji:2000id}
allowing to implement the constraints from
$P$
and
$T$
invariance is to switch to the cross channel
and match the
$J^{PC}$
quantum numbers of the corresponding tower of local operators to those of the nucleon-antinucleon state
$\langle N \bar{N}|$.
The structure of the form factor expansion and the number of the independent generalized form factors is the same in all channels related
by crossing transformation.
Therefore, due to the
$CPT$
invariance, this kind of
$J^{PC}$
matching in the cross channel automatically ensures the
$T$
invariance and the correct counting of the independent generalized form factors of
the operator matrix element in the direct channel.

In this section we summarize the findings of
Refs.~\cite{Hagler:2004yt,Ji:2000id}
and write down the system of
$J^{PC}$
selection rules for the
$t$-channel exchanges
contributing into the Mellin moments of leading of twist quark and gluon helicity flip GPDs. The explicit expressions for the parametrization
of the nucleon matrix elements of the relevant towers of quark and gluon helicity flip operators rewritten in terms of the notations consistent
with the set of conventions of
Sec.~\ref{Sec_Preliminaries}
are for convenience presented in the
Appendix~\ref{App_A}.

\subsection{Quantum number selection rules: quark case}

For the case of quark helicity flip GPDs
one has to consider the tower of local operators
\be
\hat{O}_{ T}^{q\, \mu \nu, \, \mu_1  \ldots \mu_N}= {\mathbb{S}  }_{\{ \nu \mu_1...\mu_N\}} \bar{\Psi}(0)
i \sigma^{\mu \nu} (\overleftrightarrow{i D}_{\mu_1})...(\overleftrightarrow{i D}_{\mu_N}) \Psi(0).
\label{tower_loc_quark}
\ee
As pointed out in
\cite{Hagler:2004yt},
the tower of local operators
(\ref{tower_loc_quark})
transforms according to
$\left( \frac{N+2}{2}, \frac{N}{2} \right) \oplus \left( \frac{N}{2}, \frac{N+2}{2} \right) $
representation of the Lorentz group.  Its charge conjugation parity is given by
$C=(-1)^{N+1}$.
On the contrary,
$P$
parity is not uniquely defined.
{\it E.g.} consider $t^{\mu \nu}=\bar{\Psi} i \sigma^{\mu \nu } \Psi$.
One can check that
$t^{0i}$
and
$t^{ik}$
(where as usual the Latin indices refer to the spatial directions) have different parity, $P=-1$ and $P=+1$ respectively.
In particular, this means that
$t^{+i}$
has no definite
$P$
parity. Therefore, contrary to the case of vector and pseudovector towers of local operators (which have definite
$P$
parity), both the natural
($P=(-1)^J$)
and unnatural
($P=(-1)^{J+1}$)
parity
$t$-channel
exchanges may contribute to the generalized form factors occurring in the parametrization
(\ref{FF_dec_Haegler})
of the nucleon matrix element of
(\ref{tower_loc_quark}).
This leads to two sequences of possible quantum numbers of resonances
\cite{Hagler:2004yt,Ji:2000id}:
\begin{itemize}
\item Of natural parity $P=(-1)^J$:
\be
J^{PC}=J^{(-)^{J} (-)^{N+1}}= 1^{- (-)^{N+1}}, \; 2^{+ (-)^{N+1}},  \; 3^{+ (-)^{N+1}}, \ldots {(N+1)}^{(-)^{N+1}  (-)^{N+1}}.
\label{NPseries quarks}
\ee
\item Of unnatural parity $P=(-1)^{J+1}$:
\be
J^{PC}=J^{(-)^{J+1} (-)^{N+1}}= 1^{+ (-)^{N+1}}, \; 2^{- (-)^{N+1}},  \; 3^{+ (-)^{N+1}}, \ldots {(N+1)}^{(-)^{N+2}  (-)^{N+1}}.
\label{ANPseries quarks}
\ee
\end{itemize}

Now, following  the analysis of
\cite{Ji:2000id},
we list below the possible
$\langle N \bar{N}|$
states with the total angular momentum
$J$.
Let
$S$
be the spin
($S=0,1$)
and
$L$
be the orbital angular momentum of the
$N \bar{N}$
system:
 $\vec{J}=\vec{L}+\vec{S}$.
The
$P$
parity of
$\langle N \bar{N}|$
state is given by
$P=(-1)^{L+1}$;
The
$C$
parity is given by
$C=(-1)^{L+S}$.
The list of allowed
$J^{PC}_L$
of
$\langle N \bar{N}|$
states then reads
\cite{Ji:2000id}:
\be
&&
0^{++}_1, \ \ \ 0^{-+}_0, \nonumber \\ &&
1^{++}_1, \ \ \ 1^{+-}_1, \ \ \ 1^{--}_0, \ \ \ 1^{--}_2;  \nonumber \\ &&
2^{++}_1, \ \ \ 2^{++}_3, \ \ \ 2^{-+}_2, \ \ \ 2^{--}_2;  \nonumber \\ &&
3^{++}_3, \ \ \ 3^{+-}_3, \ \ \ 3^{--}_2, \ \ \ 3^{--}_4;  \nonumber \\ &&
\ldots
\label{allowedPP}
\ee

Matching
(\ref{NPseries quarks}), (\ref{ANPseries quarks})
with the set of the allowed
$\langle N \bar{N}|$
states
(\ref{allowedPP})
gives
\cite{Hagler:2004yt,Ji:2000id}
the two sets of the $t$-channel spin
$J$
exchanges contributing into the Mellin moments of quark helicity flip GPDs.
They are summarizes in
Tables~\ref{Table_NP_quarks}, \ref{Table_ANP_quarks}.
\begin{table}[H]
\begin{center}
\caption{ Natural parity
$J^{PC}_L$
$t$-channel
exchanges contributing into the Mellin moments of quark helicity flip GPDs.
}
\begin{tabular}{|l|l|l|l|l|l|l|}
\hline
$N \backslash J$ & $1$ & $2$ & $3$& $4$ & $\ldots$ & Number of FFs  \\
\hline
$0$  &  $1^{--}_{0,2} $&   &  &  & & 2\\
\hline
$1$  &   & $2^{++}_{1,3}$ &  &   & & 2\\
\hline
$2$  & $1^{--}_{0,2}$  &   & $3^{--}_{2,4}$ &   & & 4\\
\hline
$3$  &   & $2^{++}_{1,3}$  &  & $4^{++}_{3,5}$  & & 4\\
\hline
\end{tabular}
\label{Table_NP_quarks}
\end{center}
\end{table}

\begin{table}[H]
\begin{center}
\caption{ Unnatural parity
$J^{PC}_L$
$t$-channel
exchanges contributing into the Mellin moments of quark helicity flip GPDs.}
\begin{tabular}{|l|l|l|l|l|l|l|}
\hline
$N \backslash J$ & $1$ & $2$ & $3$& $4$ & $\ldots$ & Number of FFs  \\
\hline
$0$  &  $1^{+-}_{1} $&   &  &  & & 1\\
\hline
$1$  &       $1^{++}_1$  & $2^{-+}_{2}$ &  &   & & 2\\
\hline
$2$  & $1^{+-}_{1}$  &  $2^{--}_{2}$ & $3^{+-}_{3}$ &   & & 3\\
\hline
$3$  &  $1^{++}_{1}$ & $2^{-+}_{2}$  & $3^{++}_{3}$ & $4^{-+}_{4}$  & & 4\\
\hline
\end{tabular}
\label{Table_ANP_quarks}
\end{center}
\end{table}

The column ``Number of form factors (FFs)''
shows the number of the generalized form factors in the parametrization
(\ref{FF_dec_Haegler})
populated by the corresponding
$J^{PC}_L$ $t$-channel
exchanges for the given
$N$.

\subsection{Quantum number selection rules: gluon case}
\label{SubSec_QN_gluons}

To consider the case of the gluon helicity flip GPDs
one has to introduce the tower of local gluon operators
\be
\hat{O}_{  T}^{g \, \alpha \rho \, \beta \sigma \, \mu_1  \ldots \mu_N}= { \mathbb{S}}_{\{ \alpha \beta \mu_1...\mu_N\}} G^{\alpha \rho}(0)   
(\overleftrightarrow{i D}_{\mu_1})...(\overleftrightarrow{i D}_{\mu_N}) G^{\beta \sigma}(0).
\label{tower_loc_gluon}
\ee
The Mellin moments of gluon helicity flip GPDs are given by the nucleon matrix elements of the operators
(\ref{tower_loc_gluon})
convoluted with the usual projection operator
\be
\tau_{i j; \, \rho \sigma}^\bot n^\alpha n^\beta.
\ee
The tower of the local operators
(\ref{tower_loc_gluon})
transforms according to
$\left(\frac{N+4}{2}, \frac{N}{2} \right) \oplus \left(\frac{N+2}{2}, \frac{N+2}{2} \right) \oplus \left(\frac{N}{2}, \frac{N+4}{2} \right)$
representation of the Lorentz group.
The explicit expression for the parametrization of the nucleon matrix elements of the operators
(\ref{tower_loc_gluon})
is given in
(\ref{FF_dec_My}).
Its charge conjugation parity is given by
$C=(-1)^{N+2}$. Note, that as the gluon is its own antiparticle the gluon helicity flip GPDs
$H^g_T$,
$\tilde{H}^g_T$,
$E^g_T$
and
$\tilde{E}^g_T$
are even functions of
$x$.
Therefore we need to consider only even Mellin moment of the gluon helicity flip GPDs%
\footnote{
At this point we disagree with the analysis of 
Ref.~\cite{Chen:2004cg} in which
odd Mellin moments of gluon helicity flip GPDs are also considered. See discussion in Sec.~3.2 . }

Similarly to the quark case, the tower of operators
(\ref{tower_loc_gluon})
does not  possess the  definite $P$ parity.
This leads to two sequences of possible quantum numbers of the $t$-channel exchanges:
\begin{itemize}
\item Of natural parity $P=(-1)^J$:
\be
J^{PC}=J^{(-)^{J} (-)^{N+2}}= 1^{- (-)^{N+2}}, \; 2^{+ (-)^{N+2}},  \; 3^{+ (-)^{N+2}}, \ldots {(N+2)}^{(-)^{N+2}  (-)^{N+2}}.
\ee
\item Of unnatural parity $P=(-1)^{J+1}$:
\be
J^{PC}=J^{(-)^{J+1} (-)^{N+2}}= 1^{+ (-)^{N+2}}, \; 2^{- (-)^{N+2}},  \; 3^{+ (-)^{N+2}}, \ldots {(N+2)}^{(-)^{N+2}  (-)^{N+1}}.
\ee
\end{itemize}
Now  matching these two sequences with the set of the allowed
$\langle N \bar{N}|$
states
(\ref{allowedPP})
one can work out the selection rules
\cite{Chen:2004cg}
for the quantum numbers of the
$t$-channel
exchanges contributing into the Mellin moments of gluon helicity flip GPDs.
The results for even $N$ are summarized in the
Tables~\ref{Table_NP_gluons}, \ref{Table_ANP_gluons}.

\begin{table}[H]
\begin{center}
\caption{ Natural parity $J^{PC}_L$ $t$-channel
exchanges contributing into the even Mellin moments of gluon helicity flip GPDs. }
\begin{tabular}{|l|l|l|l|l|l|l|l|l|}
\hline
$N \backslash J$ & $1$ & $2$ & $3$& $4$ & $5$& $6$& $\ldots$ & Number of FFs  \\
\hline
$0$  &  & $2^{++}_{1,3}$  &  &  & & & &2\\
\hline
$2$  &    & $2^{++}_{1,3}$  &   & $4^{++}_{3,5}$  & & & & 4\\
\hline
$4$  &   & $2^{++}_{1,3}$  &  & $4^{++}_{3,5}$  & & $6^{++}_{5,7}$ & & 6\\
\hline
\end{tabular}
\label{Table_NP_gluons}
\end{center}
\end{table}

\begin{table}[H]
\begin{center}
\caption{ Unnatural parity $J^{PC}_L$ $t$-channel
exchanges contributing into the even Mellin moments of gluon helicity flip GPDs. }
\begin{tabular}{|l|l|l|l|l|l|l|l|l|}
\hline
$N \backslash J$ & $1$ & $2$ & $3$& $4$ & $5$& $6$& $\ldots$ & Number of FFs  \\
\hline
$0$  &  & $2^{-+}_{2}$  &  &  & & & &1\\
\hline
$2$  &    & $2^{-+}_{2}$  &  $3^{++}_3$ & $4^{-+}_{4}$  & & & & 3\\
\hline
$4$  &   & $2^{-+}_{2}$  & $3^{++}_3$ & $4^{-+}_{4}$  & $5^{++}_5$ & $6^{-+}_{6}$ & & 5\\
\hline
\end{tabular}
\label{Table_ANP_gluons}
\end{center}
\end{table}

From Tables~\ref{Table_NP_gluons}, \ref{Table_ANP_gluons}
we conclude that we inevitably need the contribution of unnatural parity mesons to
obtain the proper number of independent form factors for the $N$-th Mellin moment  of gluon helicity GPDs.

\setcounter{equation}{0}

\section{Miscellaneous}
\label{App_B}
\subsection{Conventions for ordinary helicity spinors}
Following App.~B of Ref.~\cite{Diehl} we use the following conventions
for the ordinary helicity spinors:
\be
&&
U(p,+)=   \begin{pmatrix} \sqrt{p^0+m} \, \chi_+(p)  \\  \sqrt{p^0-m}  \, \chi_+(p) \end{pmatrix}; \ \ \
 U(p,-)=   \begin{pmatrix} \sqrt{p^0+m} \, \chi_-(p)  \\  -\sqrt{p^0-m}  \, \chi_-(p) \end{pmatrix}; \nonumber \\ &&
 V(p,+)=   -\begin{pmatrix} \sqrt{p^0-m} \, \chi_-(p)  \\  -\sqrt{p^0+m}  \, \chi_-(p) \end{pmatrix}; \ \ \
 V(p,-)=   -\begin{pmatrix} \sqrt{p^0-m} \, \chi_+(p)  \\  \sqrt{p^0+m}  \, \chi_+(p) \end{pmatrix},
\label{helicity_spinors}
\ee
where the two-spinors read
\be
&&
\chi_+(p)= \frac{1}{\sqrt{2 |\vec{p}|(|\vec{p}|+p^3)}} \begin{pmatrix} |\vec{p}|+p^3  \\  p^1+i p^2 \end{pmatrix}; \nonumber \\ &&
\chi_-(p)= \frac{1}{\sqrt{2 |\vec{p}|(|\vec{p}|+p^3)}} \begin{pmatrix} -p^1+i p^2  \\  |\vec{p}|+p^3.    \end{pmatrix}.
\ee
Here $p=(p_0,\vec{p}) \equiv (p^0, p^1,p^2,p^3)$ is the corresponding momentum four-vector.

\subsection{Wigner $d$-functions}

\label{App_d_functions}
The expression of the Wigner $d$-functions through the Jacobi polynomials (see eq. (3.72) of \cite{Angular}) :
\be
d^J_{m'm}(\theta)= \sqrt{\frac{(J+m)! (J-m)!}{(J+m')! (J-m')!}}  \left( \sin \frac{\theta}{2} \right)^{m-m'}  \left( \cos \frac{\theta}{2} \right)^{m+m'}
P_{J-m}^{(m-m',\,m+m')} (\cos \theta).
\label{hvast}
\ee
The relation to the usual Wigner function is given by $d^J_{m'm}(\theta)  = D^J_{m'm}(0,\theta,0)$.

Wigner $d$-functions $d^J_{m, m'}$ are orthogonal for fixed $m$, $m'$:\be
\int_0^\pi d \theta  \, \sin \theta \, d^J_{m, m'}(\theta)  d^{J'}_{m, m'} (\theta) =
\frac{2}{2J+1} \delta_{J'J}.
\ee

The relation
(\ref{hvast})
together with the Rodriguez formula for the Legendre polynomials helps
to express the Wigner $d$-functions employed in the present paper through the Legendre polynomials
and their derivatives.

\end{document}